\documentclass[
  journal=pasa,
  manuscript=article-type,
  year=2020,
  volume=37,
]{cup-journal}

\usepackage{amsmath}
\usepackage[nopatch]{microtype}
\usepackage{booktabs}
\usepackage{mathtools}
\usepackage{amssymb}
\usepackage[caption=false]{subfig}
\usepackage{tabularx}
\usepackage{multirow}
\usepackage{hyperref}

\DeclareMathAlphabet\mathbfcal{OMS}{cmsy}{b}{n}

\title{Comparison of Fast, Hybrid Imaging Architectures for Multi-scale, Hierarchical Aperture Arrays}

\author{Nithyanandan Thyagarajan}
\affiliation{CSIRO, Space \& Astronomy, P. O. Box 1130, Bentley, WA 6102, Australia}
\email[Nithyanandan Thyagarajan]{Nithyanandan.Thyagarajan@csiro.au}

\keywords{Astronomical instrumentation, methods and techniques; instrumentation: interferometers; techniques: image processing; techniques: interferometric; telescopes} 

\begin{document}

\begin{abstract}
Two major areas of modern radio astronomy, namely, explosive astrophysical transient phenomena and observations of cosmological structures, are driving the design of aperture arrays towards large numbers of low-cost elements consisting of multiple spatial scales spanning the dimensions of individual elements, the size of stations (groupings of individual elements), and the spacing between stations. Such multi-scale, hierarchical aperture arrays require a combination of data processing architectures-- pre-correlation beamformer, generic version of FFT-based direct imager, post-correlation beamformer, and post-correlation FFT imager-- operating on different ranges of spatial scales to obtain optimal performance in imaging the entire field of view. Adopting a computational cost metric based on the number of floating point operations, its distribution over the dimensions of discovery space, namely, field of view, angular resolution, polarisation, frequency, and time is examined to determine the most efficient hybrid architectures over the parameter space of hierarchical aperture array layouts. Nominal parameters of specific upcoming and planned arrays-- the SKA at low frequencies (\texttt{SKA-low}), \texttt{SKA-low-core}, a proposed long baseline extension to \texttt{SKA-low} (\texttt{LAMBDA-I}), compact all-sky phased array (\texttt{CASPA}), and a lunar array (\texttt{FarView-core})-- are used to determine the most optimal architecture hierarchy for each from a computational standpoint, and provide a guide for designing hybrid architectures for multi-scale aperture arrays. For large, dense-packed layouts, a FFT-based direct imager is most efficient for most cadence intervals, and for other layouts that have relatively lesser number of elements or greater sparsity in distribution, the best architecture is more sensitive to the cadence interval, which in turn is determined by the science goals.
\end{abstract}

\section{Introduction}

The variable and transient radio sky encompasses a vast range of astrophysical phenomena. Notable examples include pulsars \citep{Keane2013}, fast radio bursts \citep[FRB;][]{Lorimer+2007,Thornton+2013}, magnetars and their potential link to FRBs \citep{Bochenek+2020}, rotating radio transients \citep[RRAT][]{Mclaughlin+2006}, gamma-ray burst (GRB) afterglows, Jovian and solar bursts, flare stars, cataclysmic variables (CVs), exoplanet and stellar outbursts \citep{Zhang+2023}, X-ray binaries, novae, supernovae, active galactic nuclei (AGNs), blazars, tidal disruption events, counterparts to gravitational wave (GW) events, and several phenomena yet to be classified \citep{Thyagarajan+2011}. Their timescales span orders of magnitude, from nanoseconds to days, across wide frequency ranges and polarisation states \citep[][and references therein]{Pietka+2015,Chandra+2016,Nimmo+2022}. For example, pulsars and their giant radio pulses can operate on timescales from milliseconds to seconds, with millisecond pulsars on milliseconds to tens of milliseconds, while FRBs are active on sub-millisecond to tens of millisecond timescales \citep{Crawford+2022,Gupta+2022,Snelders+2023}. Extremely energetic and ultra-fast phenomena, such as sub-millisecond pulsars and magnetars \citep{Du+2009,Haskell+2018}, and nanosecond-scale giant pulses from the Crab pulsar \citep{Hankins+2003,Hankins+2007,Eilek+2016,Philippov+2019}, can reveal the fundamental nature of matter and exotic physics. The effectiveness of discovering these transients depends on the instrument's field of view and the duration of on-sky observation \citep{Cordes2007}, ideally requiring continuous real-time operation over a large instantaneous field of view to capture phenomena ranging from microseconds to days.

Another significant goal of modern radio astronomy is the exploration of structure formation and constraining of cosmological parameters in the early Universe using the redshifted 21~cm line of neutral Hydrogen as a tracer of large-scale structures \citep[][and references therein]{Morales+2010,Pritchard+2012}. One prominent example includes probing the Cosmic Dawn and the Epoch of Reionisation signifying the appearance of the first luminous objects in the Universe and their impact on structure formation in the Universe at $z\gtrsim 6$. Another example is to probe the Dark Energy and acceleration of the Universe at $1\lesssim z\lesssim 4$ using intensity mapping of redshifted 21~cm from H{\sc i} bound to galaxies \citep{CosmicVisions+2018}. Such cosmological radio observations probing large-scale structures through spatial power spectrum or imaging require interferometer arrays of large fields of view (hundreds to thousands of square degrees), dense sampling of the large-scale spatial modes (tens of kpc to tens of Mpc), and high sensitivity ($\lesssim 1$~mK noise levels). 

Based on the requirements for transients and observational cosmology, radio interferometers are witnessing a paradigm shift towards packing a large number of relatively small-sized collecting elements densely in a compact region. Modern radio arrays tend to be aperture arrays that operate interferometrically and hierarchically on multiple spatial scales, involving spatial dimensions of the collecting element, a spatial grouping of elements forming a virtual telescope typically referred to as a station or a tile, and a spatial grouping of stations comprising the full array.
Examples of multi-scale, hierarchical aperture arrays include the Murchison Widefield Array \cite[MWA;][]{Tingay+2013}, the Hydrogen Epoch of Reionization Array \citep[HERA;][]{HERA+2017}, the Low Frequency Array \cite[LOFAR;][]{vanHaarlem+2013}, the swarm of Long Wavelength Arrays \cite[LWA Swarm;][]{Dowell+2018}, 
the Hydrogen Intensity and Real-time Analysis Experiment \citep[HIRAX;][]{HIRAX+2022}, and the \texttt{SKA-low} \citep{Dewdney+2009,SKA1+2019}. 

This new paradigm poses severe computational and data rate challenges necessitated by the processing of large numbers of data streams in short cadence intervals. The choice of data processing architecture is usually sensitive to the array layout, cadence, angular resolution, and field of view coverage. For example, a correlator-based architecture is generally more suited for relatively smaller number of interferometer elements sparsely distributed over a large area and processing on a slower cadence, whereas a direct imaging architecture based on the Fast Fourier Transform (FFT) has far greater efficiency for regularly arranged elements \citep{Daishido+1991,Otobe+1994,Tegmark+2009,Tegmark+2010,Foster+2014,Masui+2019}. A generalised version of FFT-based direct imaging like the E-field Parallel Imaging ``Correlator'' \citep[EPIC;][]{Thyagarajan+2017,Thyagarajan+2019,Krishnan+2023} is expected to be highly efficient for large-$N$, densely packed arrays with or without regularly placed elements from computational and data rate perspectives. Pre- or post-correlation beamforming is expected to be efficient when the elements and the image locations are few in number. Thus, the choice of an optimal imaging architecture is critical for maximising the discovery potential. If the chosen architecture is inadequate, one or more discovery parameters such as time resolution, on-sky observing time, bandwidth, field of view, or angular resolution must be curtailed to obtain a compromised optimum \citep[for example,][]{Price2024}. For multi-scale arrays that involve hierarchical data processing, a single architecture may not be optimal on all scales. A hybrid architecture that is optimal at various levels of data processing hierarchy is required. 

In this paper, the primary motivation is to explore different imaging architectures and their combinations appropriate for aperture arrays spanning a multi-dimensional space of array layout parameters and cadence intervals from a computational viewpoint. Conversely, the paper also provides a guide to choosing the array layout parameters and cadence intervals where a given architecture, or combinations thereof, will remain computationally efficient. Several upcoming and hypothetical multi-scale aperture arrays are used as examples. 

The paper is organised as follows.
Section~\ref{sec:multi-scale-arrays} describes the parametrisation of multi-scale aperture arrays studied. Section~\ref{sec:computational-cost} describes the metric of computational cost density in discovery phase space based on which different imaging architectures described in section~\ref{sec:img-archs} are compared. The multi-scale, intra- and inter-station imaging architecture options are described in sections~\ref{sec:intra-station-arch} and \ref{sec:inter-station-arch}, respectively. Section~\ref{sec:results} contains a discussion and summary of the results, and conclusions are presented in section~\ref{sec:conclusion}. 

\section{Multi-scale, Hierarchical Aperture Arrays} \label{sec:multi-scale-arrays}

\begin{figure}
\includegraphics[width=\linewidth]
{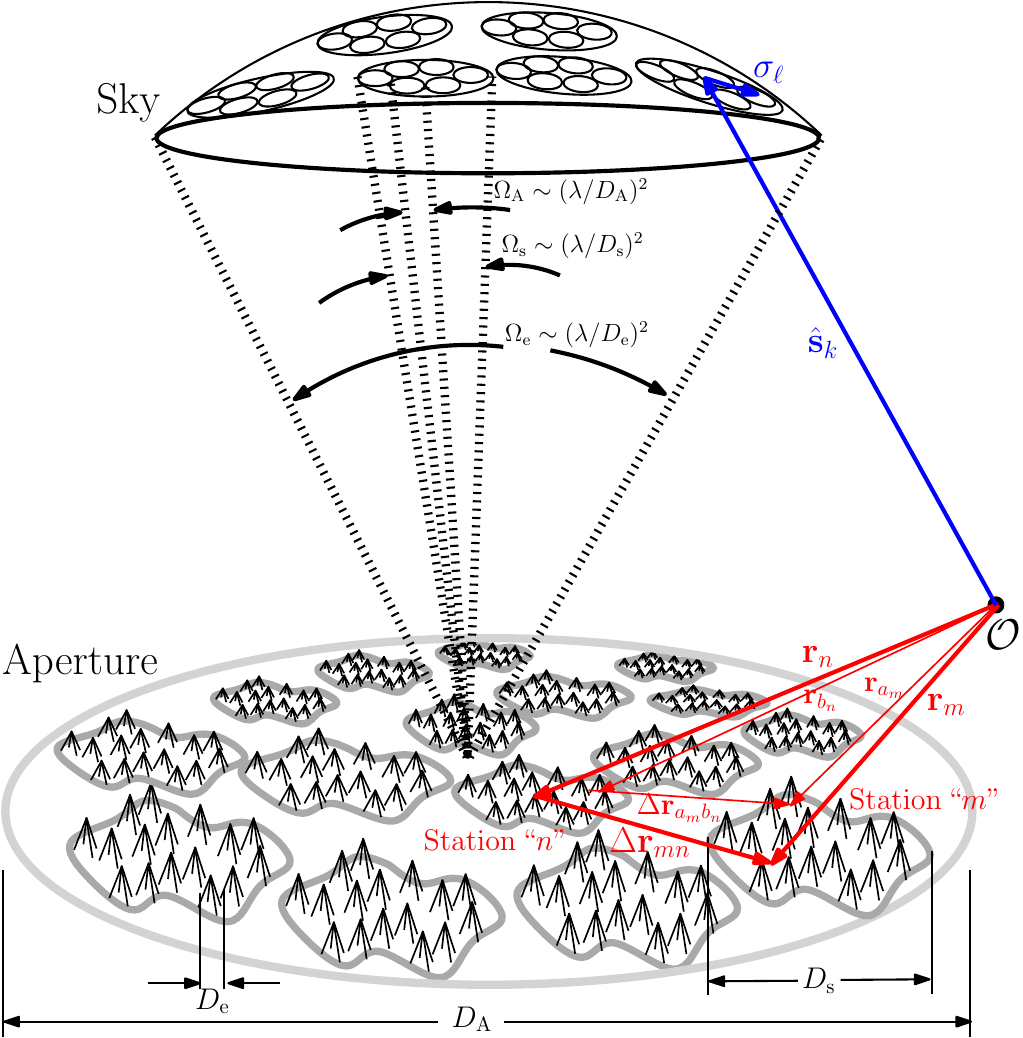}
\caption{Geometric view of a hierarchical multi-scale aperture array, which consists of $N_\textrm{s}$ stations, each a collection of $N_\textrm{eps}$ elements. Stations, $m$ and $n$, at locations, $\boldsymbol{r}_m$ and $\boldsymbol{r}_n$, respectively, relative to the origin, $\mathcal{O}$, are separated by $\Delta\boldsymbol{r}_{mn}$. Elements, $a$ and $b$, in stations, $m$ and $n$, are denoted by $a_m$ and $b_n$, located at $\boldsymbol{r}_{a_m}$ and $\boldsymbol{r}_{b_n}$, respectively. $\Delta\boldsymbol{r}_{a_m b_n}$ is the separation between these elements. The spatial sizes of the element, station, and the array in the aperture plane are denoted by $D_\textrm{e}$, $D_\textrm{s}$, and $D_\textrm{A}$, respectively. The solid angle, $\Omega_\textrm{e}\sim (\lambda/D_\textrm{e})^2$, corresponding to the element size determines the overall field of view in the sky plane accessible by the aperture array. Intra-station architectures produce station-sized virtual telescopes that simultaneously produce beams of solid angle, $\Omega_\textrm{s}\sim (\lambda/D_\textrm{s})^2$, in multiple directions, $\hat{\boldsymbol{s}}_k$, that can fill $\Omega_\textrm{e}$. Inter-station processing towards any of these beams produces virtual telescopes of size, $D_\textrm{A}$, that simultaneously produce beams of solid angle, $\Omega_\textrm{A}\sim (\lambda/D_\textrm{A})^2$, in multiple directions, $\boldsymbol{\sigma}_\ell$, relative to $\hat{\boldsymbol{s}}_k$, that can fill the solid angle of that beam. 
\label{fig:geometry-notation}}
\end{figure}

The hierarchical, multi-scale aperture array scenario considered involves spatial scales corresponding to the dimensions of the collecting element, station (a grouping of elements), and the array (a grouping of stations), denoted by $D_\textrm{e}$, $D_\textrm{s}$, and $D_\textrm{A}$, respectively. Figure~\ref{fig:geometry-notation} illustrates the geometry and notation for hierarchical multi-scale aperture arrays studied in this paper.

The number of stations in the array and the number of elements per station (indexed by $m$) are denoted by $N_\textrm{s}$ and $N_{\textrm{eps}_m}$, respectively. The number of elements in all stations are assumed to be equal, $N_{\textrm{eps}_m} = N_\textrm{eps}$, for convenience. Without loss of generality, this study can be extended to arrays with heterogeneous station densities and layouts such as LOFAR, but the idiosyncrasies of the array will have to be taken into account.

The specific aperture array examples considered are the \texttt{SKA-low}, the core of the \texttt{SKA-low} (\texttt{SKA-low-core}), the first phase of the concept of a long baseline extension to \texttt{SKA-low} called Low-frequency Australian Megametre Baseline Demonstrator Array (\texttt{LAMBDA-I}\footnote{\url{https://www.atnf.csiro.au/projects/LAMBDA.html}}), the Compact All-Sky Phased Array \citep[\texttt{CASPA}\footnote{Presently, CASPA's stations are intended for localisation and astrometry, rather than for aperture synthesis.};][]{Luo+2024}, and the core of a lunar array \citep[\texttt{FarView};][]{Polidan+2024} which is hereafter referred to as \texttt{FarView-core}. \texttt{LAMBDA-I}, \texttt{CASPA}, and \texttt{FarView-core} are hypothetical interferometers that are still in concept phase.

The parameters for the chosen examples are summarised in Table~\ref{tab:array_params}. Nominal wavelengths, $\lambda$, at which the arrays are planned to operate are listed. The computational costs are typically independent of the wavelength, and only depend on the spatial frequencies as shown in Table~\ref{tab:array_params}. However, the aperture efficiencies can, in general, vary with wavelength. For a fair comparison, the apertures are all assumed to be 100\% efficient at the nominal wavelengths chosen, and that the arrays are used for real-time imaging.

\begin{table*}[htb!]
\normalsize
\begin{threeparttable}
\caption{Table of hierarchical, multi-scale interferometer array parameters.}
\label{tab:array_params}
\begin{tabular}{ccccccc|cc}
\toprule
\headrow Array name & $\lambda$ (m)\tnote{a} & $D_\textrm{A}/\lambda$ & $N_\textrm{s}$ & $D_\textrm{s}/\lambda$ & $N_\textrm{eps}$ & $D_\textrm{e}/\lambda$ & $f_\textrm{A}$\tnote{b} & $f_\textrm{s}$\tnote{c} \\
\midrule
\texttt{LAMBDA-I} & $2$ & $3.9\times 10^6$ & $4$ & $17.5$ & $256$ & $1$ &  $1.61\times 10^{-10}$ & $0.836$ \\ 
\texttt{SKA-low-core} & $2$ & $5.25\times 10^{2}$ & $256$ & $17.5$ & $256$ & $1$ & $0.284$ & $0.836$ \\ 
\texttt{SKA-low} & $2$ & $3.5\times 10^4$ & $512$ & $17.5$ & $256$ & $1$ &  $1.28\times 10^{-4}$ & $0.836$ \\ 
\texttt{CASPA} & $0.25$ & $4.6\times 10^4$ & $3$ & $8.08$ & $65$ & $0.5$ &  $9.2\times 10^{-8}$ & $0.996$ \\ 
\texttt{FarView-core}\tnote{d} & $10$ & $6.77\times 10^2$ & $81$ & $38.5$ & $625$ & $1$ & $0.262$ & $0.422$ \\ 
\bottomrule
\end{tabular}
\begin{tablenotes}[hang]
\item[a]Nominal operating wavelength of the array. Results may differ at other wavelengths depending on aperture efficiency, filling factor, etc. 
\item[b]Array filling factor, $\,f_\textrm{A}=N_\textrm{s}\,(D_\textrm{s}/D_\textrm{A})^2$
\item[c]Station filling factor, $\,f_\textrm{s}=N_\textrm{eps}\,(D_\textrm{e}/D_\textrm{s})^2$
\item[d]Only the 6~km core of the proposed array \citep{Polidan+2024} is considered here.
\end{tablenotes}
\end{threeparttable}
\end{table*}

Note that, traditionally, cosmological applications such as probing the epoch of reionisation (EoR) using redshifted 21~cm have relied on storing the visibilities and processing them offline. In such cases, although imaging costs like gridding, Discrete Fourier Transform (DFT), or FFT will not apply, the correlations must still be obtained in real-time. With large arrays, the traditional correlator approach may be too expensive. For example, the Packed Ultra-wideband Mapping Array \citep[PUMA;][]{PUMA+2019}, a planned radio telescope for cosmology and transients with $\sim 5000-32000$ elements, is considering an FFT correlator architecture, a specific adaptation of the EPIC architecture for redundant arrays. This study, with a breakdown of costs for different operations, is also relevant for such cosmological arrays.

\section{Cost Metric} \label{sec:computational-cost}

The primary metric for computational cost used in this study is the number of real floating point operations (FLOP) and the density of such operations in the discovery phase space volume whose dimensions span time, frequency, polarisation, and location on the sky. The computational cost will be estimated per independent interval of time (in FLOP per second or FLOPS) for all possible polarisation states per frequency channel ($\delta\nu$) per independent angular resolution element ($\delta\Omega$). In this paper, a discovery phase space volume element comprising of a single frequency channel at a single two-dimensional angular resolution element (independent pixel) will be referred to as a \textit{voxel} ($\delta\nu\,\delta\Omega$).

The cost estimates rely on the following floating point operations. A complex addition consists of two FLOPs, and a complex multiplication consists of four real multiplications and two real additions and therefore six FLOPs. A complex multiply-accumulate (CMAC) operation includes one complex multiplication and addition, therefore requiring eight FLOPs.

Besides the computational metric employed in this paper, a practical implementation of any of these architectures requires careful consideration of several other practical factors such as memory bandwidth, input and output data rates, power consumption, and other processing costs such as calibration. These factors, elaborated further in section~\ref{sec:other-metrics}, will vary significantly with real-life implementations and hardware capabilities available during deployment. Therefore, in a complex space of multi-dimensional factors, this work adopts a simple theoretical approach as a first step wherein the budget of floating point operations will form an irreducible requirement regardless of any particular implementation. 

\section{Fast Imaging Architectures} \label{sec:img-archs}

\begin{figure}
\includegraphics[width=0.95\linewidth]
{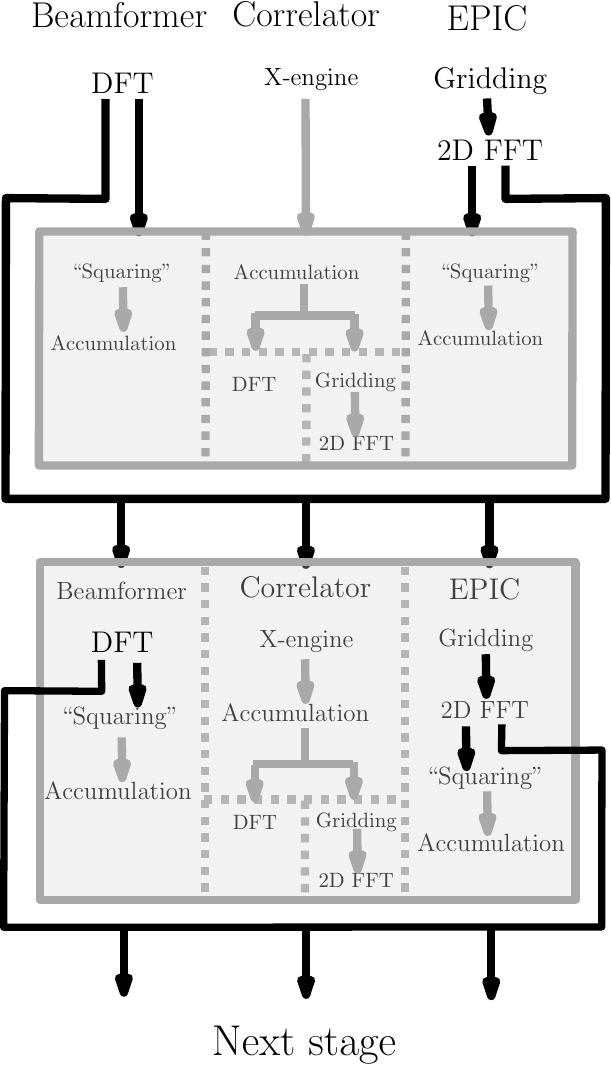}
\caption{A hybrid, two-stage architecture for hierarchical aperture arrays showing the methodology for extension to multiple stages. Black arrows indicate pathways where the signal consists of electric fields with phase coherence. Gray arrows indicate pathways of a ``squared'' or a correlated signal with no absolute but only relative phase information. At the top are intra-station architectures, namely, BF, correlator, and EPIC. The top box denotes intra-station imaging options. The bottom box denotes inter-station imaging architectures that process coherent outputs obtained from multiple stations deploying intra-station BF and/or EPIC architectures. The electric fields phase coherently synthesised by BF or EPIC at any stage can form the input for the next stage, thereby allowing the extension of the hierarchy.
\label{fig:hybrid-architectures-cost-schematic}}
\end{figure}

Two broad classes of two-stage fast imaging architectures are considered. In both classes, the first stage is intra-station interferometry of the element electric fields using the options of voltage beamforming (BF), E-field Parallel Imaging Correlator (EPIC), beamforming of correlations (XBF), or correlation followed by FFT (XFFT). The second stage involves inter-station combination of the data products from the first stage. The main difference between the two classes is in whether the second stage uses the first stage products coherently or incoherently. Although aperture arrays consisting of only three hierarchical scales are considered here, the analysis can be extended without loss of generality to arrays of even larger levels of spatial hierarchy. A schematic of these hybrid architectures, their computational cost components, and their extension methodology to multiple stages are illustrated in Figure~\ref{fig:hybrid-architectures-cost-schematic}.

Fast imaging involves processing data on multiple time scales. The channelisation of a voltage time stream of length $\delta t$ with a time resolution of $t_\textrm{s}$ will correspond to spectral channels of width $\delta\nu=1/\delta t$ and a bandwidth of $B=1/t_\textrm{s}$. The number of spectral channels is $N_\nu=B/\delta\nu=\delta t/t_\textrm{s}$. Images are expected to be produced at a cadence of $t_\textrm{acc}$, which is typically larger than $\delta t$ and smaller than the time scale on which calibration needs to be updated. In this paper, the nominal value for $\delta t$ is 25~$\mu$s, and the possibilities for $t_\textrm{acc}$ explored range from 100~$\mu$s to 10~s. 

In the discussions and equations that follow, a real-time calibration is assumed to have been performed on the input voltages. By assuming the system is reasonably stable, a much less frequent ($\ll 1/t_\textrm{acc}$) calibration will be sufficient, which therefore will not significantly affect the total cost budget \citep{Beardsley+2017,Gorthi+2021}. Although in principle an accurate calibration is possible in real time, applications such as EoR that require high-accuracy calibration have, for practical reasons, relied on offline calibration.

\section{Intra-station Imaging Architectures} \label{sec:intra-station-arch}

This first stage essentially synthesises station-sized apertures with simultaneous, multiple virtual pointings. These station-sized synthesised apertures from the first stage will act as building blocks for the second stage of inter-station processing. The element, $a$, in station, $m$ is denoted by $a_m$. The element and station locations are denoted by $\boldsymbol{r}_{a_m}$ and $\boldsymbol{r}_m$, respectively. Similarly, $\Delta\boldsymbol{r}_{a_m b_n}$ is the separation between an elements, $a$ and $b$, in stations, $m$ and $n$, respectively. $\Delta\boldsymbol{r}_{mn}$ denotes the station separation. This notation is illustrated in Figure~\ref{fig:geometry-notation}. A summary of the intra-station costs are provided in Table~\ref{tab:intra-inter-station-coherent-imaging}. 

\subsection{Voltage Beamforming (BF)}

Let $\widetilde{E}_{a_m}^{p}(\nu)$ represent the calibrated and noise-weighted electric field at frequency, $\nu$, in polarisation state, $p$, measured by a station element, $a_m$, in station, $m$. Because each spectral channel is assumed to be processed independently, specifying the frequency can be dropped for convenience to write $\widetilde{E}_{a_m}^{p}(\nu)\equiv \widetilde{E}_{a_m}^{p}$. Then, such electric fields from individual station elements
can be phase-coherently superposed to obtain the electric field in polarisation state, $\alpha$, towards any desired direction as
\begin{align}
    \widetilde{\mathcal{E}}_m^\alpha(\hat{\boldsymbol{s}}_k) &= \sum_{a_m} \sum_p  \widetilde{\mathcal{W}}_{a_m}^{\alpha p *}(\hat{\boldsymbol{s}}_k) \, \widetilde{E}_{a_m}^p \, e^{i\frac{2\pi}{\lambda} \hat{\boldsymbol{s}}_k\cdot\boldsymbol{r}_{a_m}} \, , \label{eqn:intra-station-pol-hol-img-expl}
\end{align}
where, $\hat{\boldsymbol{s}}_k$ denotes the direction of the superposed beam, $k$,
and $\widetilde{\mathcal{W}}_{a_m}^{{\alpha p}^*}(\hat{\boldsymbol{s}}_k)$ denotes a complex-valued directional weighting\footnote{One can choose $\widetilde{\mathcal{W}}_{a_m}^{\alpha p}(\hat{\boldsymbol{s}}_k)=\mathcal{W}_{a_m}^{p\alpha}(\hat{\boldsymbol{s}}_k)$, the directional electric field sensitivity, if the signal-to-noise ratio is to be optimised. But it can also be generically chosen depending on the property desired of the estimator \cite[][]{Morales2011}.} with the superscript indices $\alpha, p$, representing the contribution of polarisation state, $p$ in the station element, $a_m$, to polarisation state, $\alpha$, towards direction, $\hat{\boldsymbol{s}}_k$. The * denotes complex conjugate. Equation~(\ref{eqn:intra-station-pol-hol-img-expl}) resembles a DFT of the calibrated Electric field measurement after applying a complex-valued weighting.

The polarised intensity in the beamformed pixel is then obtained by
\begin{align}
    \widetilde{\mathcal{I}}^{\alpha\beta}_m(\hat{\boldsymbol{s}}_k) &= \left\langle \widetilde{\mathcal{E}}_m^\alpha(\hat{\boldsymbol{s}}_k) \,  \widetilde{\mathcal{E}}_m^{\beta *}(\hat{\boldsymbol{s}}_k) \right\rangle \, , \label{eqn:intra-station-opt-pol-img-outprod}
\end{align}
where, angular brackets denote a temporal averaging across an interval of $t_\textrm{acc}$. $\widetilde{\mathcal{I}}^{\alpha\beta}_m(\hat{\boldsymbol{s}}_k)$ results from the outer product of indices $\alpha$ and $\beta$. Superscript, $\alpha\beta$, thus denotes the four pairwise combinations of polarisation states of the intensity. Even though it is an outer product over the polarisation indices, it will be referred to as a ``squaring'' operation, hereafter, for convenience because of what it reduces to if $\alpha=\beta$.

The solid angle of the beamformed pixel using the intra-station data is given by $\Omega_\textrm{s} \simeq (\lambda/D_\textrm{s})^2$. Beamforming can be applied to all independent beams ($n_\textrm{bs}$) filling the field of view with solid angle, $\Omega_\textrm{e} \simeq (\lambda/D_\textrm{e})^2$. Thus, $n_\textrm{bs} \simeq \Omega_\textrm{e}/\Omega_\textrm{s}=(D_\textrm{s}/D_\textrm{e})^2$. This beamforming step has to be executed at a cadence of $\delta t$. 

\begin{figure}
\includegraphics[width=\linewidth]
{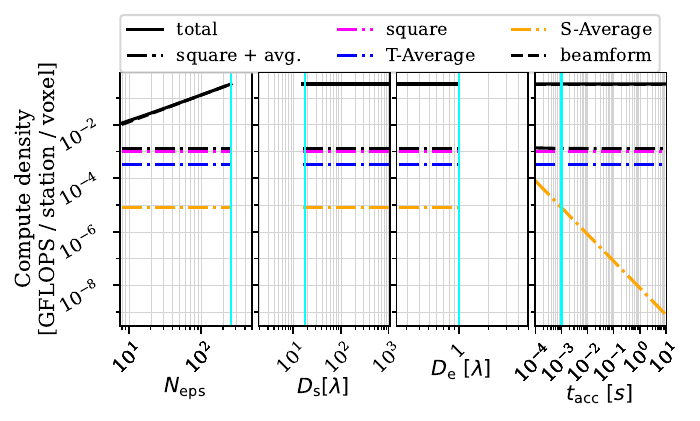}
\caption{A breakdown of the computational cost density function over the station parameters for the voltage beamforming (BF) architecture at the station level. Each panel shows the variation with the respective parameter keeping the rest fixed at the characteristic values of the \texttt{LAMBDA-I}, \texttt{SKA-low-core}, and \texttt{SKA-low} stations (cyan lines). The beamforming cost in Equation~(\ref{eqn:intra-station-pol-hol-img-expl}) denoted by black dashed lines dominates the squaring (pink dot-dashes) and temporal averaging (blue dot-dashes) costs in Equation~(\ref{eqn:intra-station-opt-pol-img-outprod}). Orange dot-dashes denote the incoherent spatial average of the intensities across stations, and has negligible contribution to the total cost.
\label{fig:1D-incoherent-compcost-BF-LAMBDA-I}}
\end{figure}

Figure~\ref{fig:1D-incoherent-compcost-BF-LAMBDA-I} shows a breakdown of the computational cost per station per voxel for voltage beamforming as a function of the different station parameters. In each panel, all parameters except the one on the $x$-axis are kept fixed at the characteristic values of the \texttt{LAMBDA-I}, \texttt{SKA-low}, and \texttt{SKA-low-core} stations as highlighted in cyan. The total cost is dominated by the DFT beamforming in Equation~(\ref{eqn:intra-station-pol-hol-img-expl}) denoted by black dashed lines. The DFT beamforming cost per voxel scales linearly with the number of elements per station. The costs of squaring and temporal averaging in Equation~(\ref{eqn:intra-station-opt-pol-img-outprod}) are subdominant. Because the squaring and temporal averaging are performed on a per-pixel basis and not dependent on $N_\textrm{eps}$, they remain constant across $N_\textrm{eps}$. And because the number of independent pixels covering the field of view scales as $n_\textrm{bs} \simeq (D_\textrm{s}/D_\textrm{e})^2$, the cost per independent pixel remains constant across $D_\textrm{s}$ and $D_\textrm{e}$. All these operations have to be performed at every time step, $\delta t$, and hence remain constant regardless of $t_\textrm{acc}$. Although spatial averaging across multiple stations (orange dot-dashed line) is not a station-level operation, it is shown to emphasise that it constitutes only a negligible cost relative to the other components, noting that it needs to be only performed on a slower cadence of $t_\textrm{acc}$ on a per-pixel basis, and thus scales inversely with $t_\textrm{acc}$.

\begin{figure}
\includegraphics[width=\linewidth]
{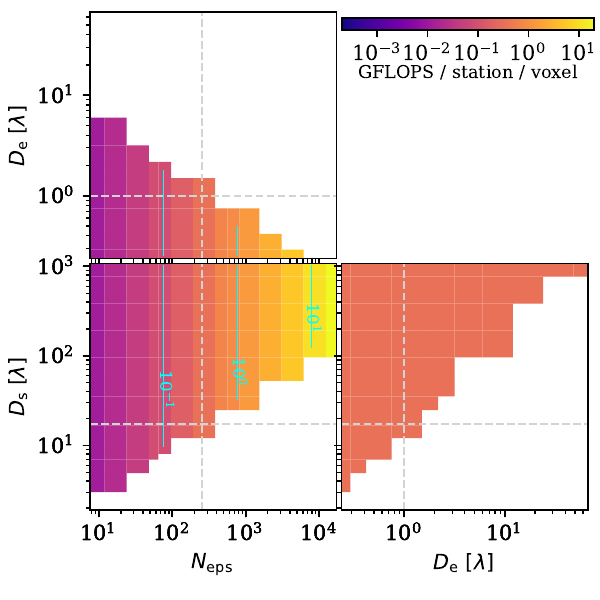}
\caption{A multi-dimensional covariant view of the net computational cost density function over the station parameters for the voltage beamforming (BF) architecture at the station level by varying two parameters at a time while fixing the rest at the nominal values of the \texttt{LAMBDA-I}, \texttt{SKA-low-core}, and \texttt{SKA-low} stations shown in grey dashed lines. The empty portions in the panels denote parts of parameter space physically impossible to sample given the constraints of the station parameters. Contours corresponding to the colour scale are shown in cyan. 
\label{fig:multidim-incoherent-compcost-BF-LAMBDA-I}}
\end{figure}

Figure~\ref{fig:multidim-incoherent-compcost-BF-LAMBDA-I} shows a covariant view of the total computational cost density per station per voxel (after summing the DFT, squaring, and averaging components) for the beamforming architecture taken two station parameters at a time while keeping the rest fixed at the nominal values of a \texttt{LAMBDA-I}, \texttt{SKA-low}, or a \texttt{SKA-low-core} station (grey dashed lines). The variation in the cost per voxel is dominated by the beamforming cost and exhibits the same trends observed in Figure~\ref{fig:1D-incoherent-compcost-BF-LAMBDA-I} -- linear in $N_\textrm{eps}$ while remaining constant with $D_\textrm{e}$ and $D_\textrm{s}$. 

\subsection{E-field Parallel Imaging Correlator (EPIC)}

Instead of DFT beamforming at discrete arbitrary locations as described above, forming independent beams that simultaneously fill the entire field of view can be achieved by a two-dimensional spatial Fast Fourier Transform (FFT) of the calibrated electric field measurements on the aperture plane. Typically, this has been associated with a regularly gridded array layout \citep{Daishido+1991,Otobe+1994,Tegmark+2009,Tegmark+2010,Foster+2014,Masui+2019}. However, it can be enabled even for an irregular layout of station elements by gridding the calibrated electric fields measured by the station elements. The Optimal Map Making \citep[OMM;][]{Tegmark1997a} formalism adopted by \citet{Morales2011} is the basis of the architecture called E-field Parallel Imaging Correlator \cite[EPIC;][]{Thyagarajan+2017},
\begin{align}
    \widetilde{\mathcal{E}}_m^\alpha(\hat{\boldsymbol{s}}_k) &= \sum_j \delta^2 \boldsymbol{r}_j \, e^{i\frac{2\pi}{\lambda} \hat{\boldsymbol{s}}_k\cdot\boldsymbol{r}_j} \left(\sum_{a_m} \sum_p \widetilde{W}_{a_m}^{\alpha p*}(\boldsymbol{r}_j-\boldsymbol{r}_{a_m}) \, \widetilde{E}_{a_m}^p \right) \, . \label{eqn:intra-station-pol-hol-img-epic}
\end{align}
The expression within the parenthesis represents the gridding operation of each calibrated electric field measurement at an arbitrary location, $\boldsymbol{r}_{a_m}$, of element $a_m$ onto a common grid at locations, $\boldsymbol{r}_j$, using a gridding kernel, $\widetilde{W}_{a_m}^{\alpha p*}(\boldsymbol{r})$ corresponding to that element. The outer summation denotes the Fourier transform of the weighted and gridded electric fields implemented through FFT. 

Application of the FFT will have the effect of simultaneously beamforming over the entire field of view, $\Omega_\textrm{e}$. The convolution with $\widetilde{W}_{a_m}^{\alpha p*}(\boldsymbol{r})$ in the aperture, which is the Fourier dual of $\widetilde{\mathcal{W}}_{a_m}^{{\alpha p}^*}(\hat{\boldsymbol{s}}_k)$ in equation~(\ref{eqn:intra-station-pol-hol-img-expl}), has several purposes. Firstly, it can be chosen to optimise specific properties in the synthesised image. For example, choosing $\widetilde{W}_{a_m}^{\alpha p*}(\boldsymbol{r})=W_{a_m}^{\alpha p*}(\boldsymbol{r})$, the complex conjugate of the holographic illumination pattern of the antenna element will optimise the signal-to-noise ratio in the image \citep{Morales2011}. Secondly, it can incorporate $w$-projection for non-coplanar element spacings \citep{Cornwell+2008}, direction-dependent ionospheric effects, wide-field distortions from refractive and scintillating atmospheric distortions, etc. \citep{Morales+2009,Morales2011}. Finally, through gridding convolution it transforms data from discrete and arbitrary locations onto a regular grid, thereby enabling the application of a two-dimensional spatial FFT. Finally, the polarised intensities are obtained by application of Equation~(\ref{eqn:intra-station-opt-pol-img-outprod}), that is pixelwise ``squaring'' (outer product of polarisation states) and temporal averaging.

\begin{figure}
\includegraphics[width=0.99\linewidth]
{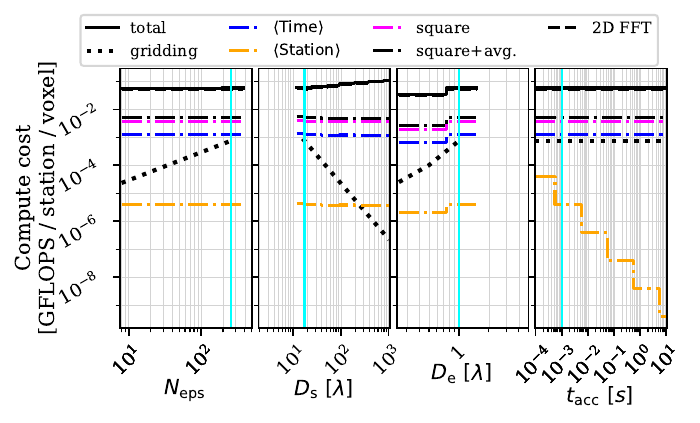}
\caption{A breakdown of the computational cost density function over the station parameters for the EPIC architecture at the station level. Each panel shows the variation with the respective parameter keeping the rest fixed at the characteristic values of the \texttt{LAMBDA-I}, \texttt{SKA-low-core}, and \texttt{SKA-low} stations (cyan lines). The spatial FFT cost in Equation~(\ref{eqn:intra-station-pol-hol-img-epic})
  denoted by black dashed lines dominates the squaring (pink dot-dashes) and temporal averaging (blue dot-dashes) costs in Equation~(\ref{eqn:intra-station-opt-pol-img-outprod}).
  The gridding operation (black dotted lines) in Equation~(\ref{eqn:intra-station-pol-hol-img-epic})
  contributes even lesser to the cost budget. The incoherent spatial  averaging of the intensities across stations (orange dot-dashed line) has negligible contribution to the total cost.
\label{fig:1D-incoherent-compcost-EPIC-LAMBDA-I}}
\end{figure}

Figure~\ref{fig:1D-incoherent-compcost-EPIC-LAMBDA-I} shows a breakdown of the costs per station per voxel of the EPIC architecture, with nominal values of a \texttt{LAMBDA-I}, \texttt{SKA-low}, or a \texttt{SKA-low-core} station marked in cyan. The dominant cost is from the two-dimensional FFT (black dashed lines) in Equation~\ref{eqn:intra-station-pol-hol-img-epic}.
The spatial FFT cost depends only on the grid dimensions. So, it is independent of $N_\textrm{eps}$ as the grid can hold as many elements as physically possible. The computational cost of the spatial FFT scales roughly as $(\gamma_\textrm{s} \, D_\textrm{s}/D_\textrm{e})^2\log_2(\gamma_\textrm{s}\, D_\textrm{s}/D_\textrm{e})^2$, where, $\gamma_\textrm{s}$ is a constant padding factor. $\gamma_\textrm{s}$ is used to control the pixel scale in the image. For example, $\gamma_\textrm{s}=2$ (adopted in this paper) along each dimension will provide identical pixel scale and image size as that formed from gridded visibilities. Notably, because the spatial FFT produces an image over the entire field of view at every independent pixel instantly, it is not a pixelwise operation, and therefore, its cost per voxel scales as $\sim \log_2(D_\textrm{s}/D_\textrm{e})$, which exhibits a weak dependence on $D_\textrm{s}$ and $D_\textrm{e}$. The squaring (pink dot-dashed lines) and temporal averaging (blue dot-dashed lines) in Equation~\ref{eqn:intra-station-opt-pol-img-outprod} operate on a per-pixel basis. Thus, those costs per voxel remain constant with $N_\textrm{eps}$, $D_\textrm{s}$ and $D_\textrm{e}$ similar to the voltage beamforming architecture. And they 
are subdominant. The gridding (black dotted lines) with weights in Equation~(\ref{eqn:intra-station-pol-hol-img-epic})
is a per-element operation and depends only on $N_\textrm{eps}$ and the size of the gridding kernel, $N_\textrm{ke}$. Here, $N_\textrm{ke}=1^2$ is chosen to provide unaliased full field-of-view images. The incorporation of $w$-projection in the gridding kernel will correspondingly increase $N_\textrm{ke}$. Nevertheless, the gridding cost, scales linearly with $N_\textrm{eps}$ and inverse squared with $D_\textrm{s}$ per voxel. Overall, the gridding cost is insignificant. Again, the spatial averaging of pixels across multiple stations (orange dot-dashed line) is not a station-level operation, but is only shown to emphasise that it operates on a much slower cadence of $t_\textrm{acc}$, scaling inversely with $t_\textrm{acc}$, and adds only negligible cost to the overall cost budget. 

\begin{figure*}
\centering
\subfloat[][Covariant cost density of EPIC architecture \label{fig:multidim-incoherent-compcost-EPIC-LAMBDA-I}]{\includegraphics[width=0.48\textwidth]
{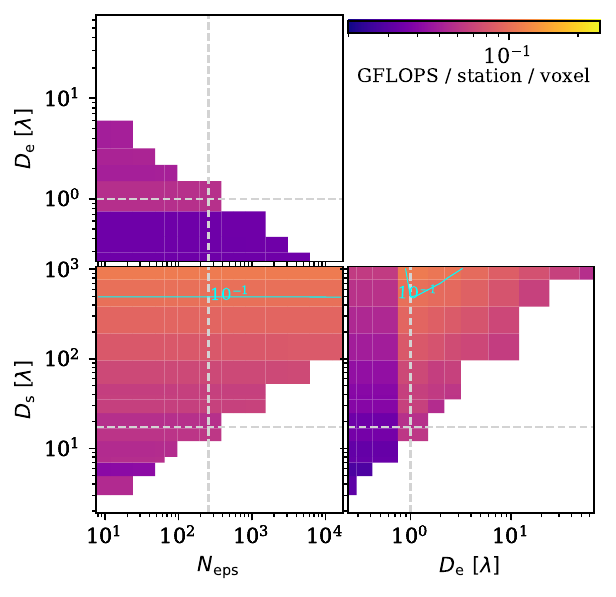}  
}
\subfloat[][Covariant cost of EPIC architecture relative to BF architecture \label{fig:multidim-incoherent-relative-compcost-EPIC-LAMBDA-I}]{\includegraphics[width=0.48\textwidth]
{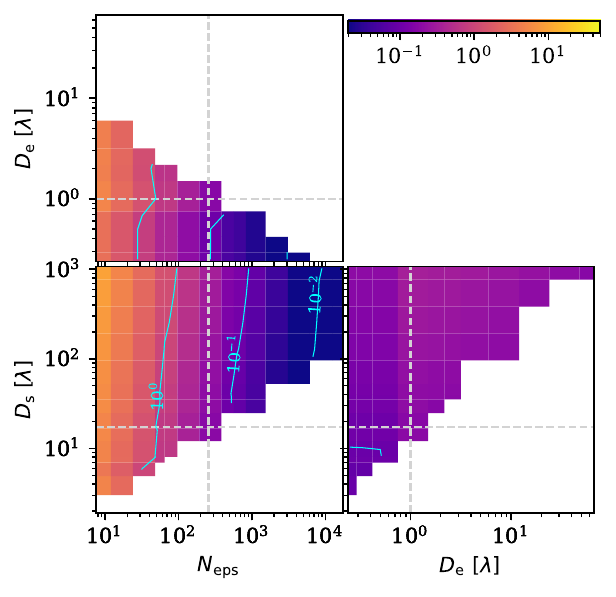}
} 
\caption{(Left): Two-dimensional slices of the computational cost per station per voxel for imaging using station-level EPIC. Because the spatial FFT dominates the overall cost budget, the computational cost density is relatively insensitive to $N_\textrm{eps}$, $D_\textrm{s}$, and $D_\textrm{e}$ as seen in Figure~\ref{fig:1D-incoherent-compcost-EPIC-LAMBDA-I}, thereby not showing much variation in the colour scale. (Right): Ratio of computational cost of EPIC to voltage beamforming. For $N_\textrm{eps}\gtrsim 100$, EPIC has a significant advantage over voltage beamforming (BF). Gray dashed lines indicate nominal values for \texttt{LAMBDA-I}, \texttt{SKA-low-core}, and \texttt{SKA-low} stations. Logarithmic contour levels corresponding to the colour scale are shown in cyan. \label{fig:incoherent-compcost-EPIC-LAMBDA-I}}
\end{figure*}

Figure~\ref{fig:multidim-incoherent-compcost-EPIC-LAMBDA-I} shows a covariant view of the total computational cost density per station per voxel (sum of gridding, FFT, squaring and temporal averaging components) for the EPIC architecture taken two station parameters at a time while keeping the rest fixed at the nominal values of a \texttt{LAMBDA-I}, \texttt{SKA-low}, or a \texttt{SKA-low-core} station denoted by grey dashed lines. The variation in the cost per voxel is dominated by the spatial FFT cost which is relatively constant in all these parameters and displays the same trends observed in Figure~\ref{fig:1D-incoherent-compcost-EPIC-LAMBDA-I}, namely, nearly constant in $N_\textrm{eps}$, $D_\textrm{e}$ and $D_\textrm{s}$. Figure~\ref{fig:multidim-incoherent-relative-compcost-EPIC-LAMBDA-I} shows the ratio of the computational cost of the EPIC architecture relative to the voltage beamforming architecture in Figure~\ref{fig:multidim-incoherent-compcost-BF-LAMBDA-I}. It is clearly noted that the cost of the EPIC architecture is significantly less than that of voltage beamforming for $N_\textrm{eps} \gtrsim 100$, and continues to decrease with increasing $N_\textrm{eps}$, implying that for full field of view imaging, EPIC has a clear and significant computational advantage over voltage DFT beamforming as $N_\textrm{eps}$ and packing density of elements in a station increase.

The EPIC architecture has been implemented on the Long Wavelength Array (LWA) station in Sevilleta (NM, USA) \citep{Kent+2019,Kent+2020,Krishnan+2023} using a GPU framework and operates on a commensal mode. With recent optimisations, the EPIC pipeline on LWA-Sevilleta computes $128\times 128$ all-sky (visible hemisphere) images at 25,000 frames per second and outputs at a cadence of 40--80~ms accumulations \citep{Reddy+2024}. For reference, the software correlator and visibility-based imager can produce images at 5~s cadence \citep{Taylor+2012}.

The EPIC architecture will be relevant for large cosmological arrays like PUMA that are considering a FFT-based correlator architecture. Obtaining visibilities from an EPIC architecture will involve an additional step of an inverse-FFT or an inverse-DFT back to the aperture plane from the accumulated images (not included in Figure~\ref{fig:1D-incoherent-compcost-EPIC-LAMBDA-I}). However, the inverse-FFT or DFT operation to obtain visibilities needs to be performed only once every accumulation interval, and is thus not expected to add significantly to the total cost budget.

\subsection{Correlator Beamforming (XBF)}

In the traditional correlator approach, a pair of electric field measurements, $E_{a_m}^p$ and $E_{b_m}^q$, in polarisation states, $p$ and $q$, at station elements, $a_m$ and $b_m$, respectively in station, $m$, can be cross-correlated and temporally averaged to obtain calibrated visibilities, $\widetilde{V}_{a_m b_m}^{pq}$. It can be written as 
\begin{align}
    \widetilde{V}_{a_m b_m}^{pq} &= \bigl\langle E_{a_m}^p \, E_{b_m}^{q*}\bigr\rangle \, . \label{eqn:intra-station-pol-visibilities}
\end{align}
The visibilities in the station can be beamformed using a DFT to obtain polarised intensities towards $\boldsymbol{s}_k$ as
\begin{align}
    \widetilde{\mathcal{I}}_m^{\alpha\beta}(\hat{\boldsymbol{s}}_k) 
    &= \sum_{a_m,b_m} \sum_{p,q} \widetilde{\mathcal{B}}_{a_m b_m}^{\alpha\beta;pq*}(\hat{\boldsymbol{s}}_k) \, \widetilde{V}_{a_m b_m}^{pq} \,  e^{i\frac{2\pi}{\lambda} \hat{\boldsymbol{s}}_k\cdot\Delta\boldsymbol{r}_{a_m b_m}} \, . \label{eqn:intra-station-pol-xbf-img-expl} 
\end{align}
$\widetilde{\mathcal{B}}_{a_m b_m}^{\alpha\beta;pq*}(\hat{\boldsymbol{s}}_k)$, is the counterpart of $\widetilde{\mathcal{W}}_{a_m}^{{p\alpha}^*}(\hat{\boldsymbol{s}}_k)$ in Equation~(\ref{eqn:intra-station-pol-hol-img-expl}) for a correlated quantity, and it denotes a complex-valued directional weighting applied to the calibrated visibility representing the contribution of polarisation state, $pq$, to a state, $\alpha\beta$, in the measurement and sky planes, respectively. It can be chosen to produce images with desired characteristics \citep{Masui+2019}. 

\begin{figure}
\includegraphics[width=\linewidth]
{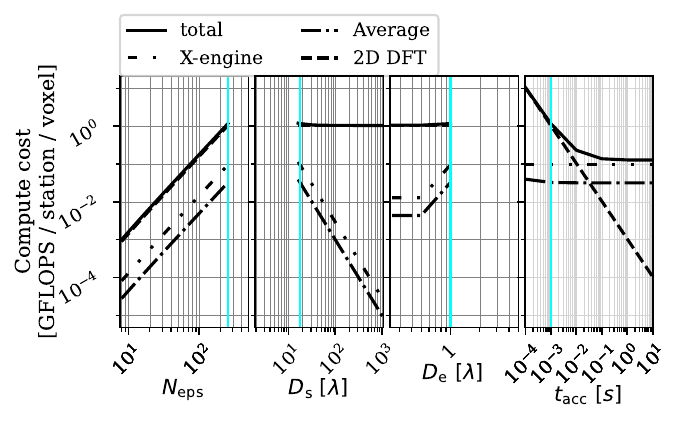}
\caption{A breakdown of the computational cost density function over the station parameters for the correlator beamforming architecture (XBF) at the station level. Each panel shows the variation with the respective parameter keeping the rest fixed at the characteristic values of the \texttt{LAMBDA-I}, \texttt{SKA-low-core}, and \texttt{SKA-low} stations (cyan lines). The two-dimensional DFT (dashed lines) dominates the cost for the chosen parameters at an imaging cadence of $t_\textrm{acc}=1$~ms over correlation / X-engine (double dot-dashed) and temporal averaging of the correlations (dot-dashed).
\label{fig:1D-incoherent-compcost-XBF-LAMBDA-I}}
\end{figure}

Figure~\ref{fig:1D-incoherent-compcost-XBF-LAMBDA-I} shows the computational cost density breakdown for correlator beamforming at the station level. The correlation and temporal averaging in Equation~(\ref{eqn:intra-station-pol-visibilities}) are performed on every pair of station elements and scale as $N_\textrm{eps}(N_\textrm{eps}-1)/2$, with no dependence on $D_\textrm{e}$ or $D_\textrm{s}$. Hence, their costs per voxel scale as $(D_\textrm{s}/D_\textrm{e})^{-2}$. The two-dimensional DFT in Equation~(\ref{eqn:intra-station-pol-xbf-img-expl}) also scales as $N_\textrm{eps}(N_\textrm{eps}-1)/2$, but being a per-pixel operation, stays constant with $D_\textrm{e}$ and $D_\textrm{s}$. The correlation and temporal averaging have to be performed at every interval of $\delta t$ and are therefore, independent of $t_\textrm{acc}$, whereas the DFT needs to be performed at a cadence of $t_\textrm{acc}$, and thus scales as $t_\textrm{acc}^{-1}$. For the chosen parameters of \texttt{LAMBDA-I}, \texttt{SKA-low}, and \texttt{SKA-low-core}, at $t_\textrm{acc}=1$~ms, the computational cost density is dominated by the cost of the DFT. However, for imaging on a slower cadence, $t_\textrm{acc}\gtrsim 10$~ms, the cost of correlation begins to dominate. 

\begin{figure*}
\centering
\subfloat[][Covariant cost density of XBF architecture \label{fig:multidim-incoherent-compcost-XBF-LAMBDA-I}]
{\includegraphics[width=0.48\textwidth]
{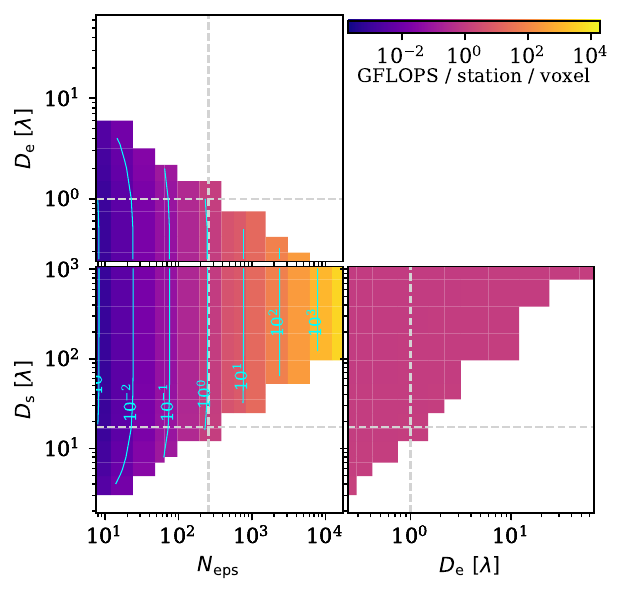}
}
\subfloat[][Covariant cost of XBF architecture relative to BF architecture \label{fig:multidim-incoherent-relative-compcost-XBF-LAMBDA-I}]
{\includegraphics[width=0.48\textwidth]
{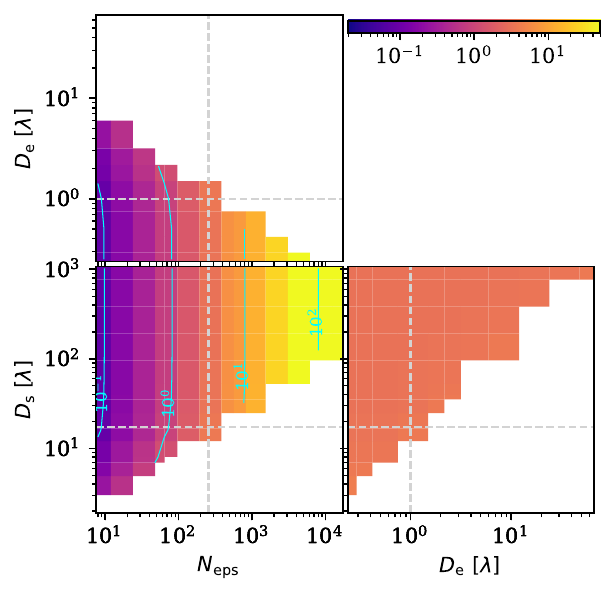}
}
\caption{(Left): Two-dimensional slices of the computational cost for imaging using a DFT beamforming of station-level cross-correlations (XBF) for \texttt{LAMBDA-I}, \texttt{SKA-low-core}, and \texttt{SKA-low} station parameters (dashed gray lines). Cyan lines denote contours of the colour scale in logarithmic increments. (Right): Same as the left but relative to a voltage beamformer (BF) architecture. The computational cost density is dominated by the two-dimensional DFT cost for $t_\textrm{acc}=1$~ms, and is lower than that of voltage beamforming (BF) for $N_\textrm{eps}\lesssim 100$ while getting more expensive with increasing $N_\textrm{eps}$. \label{fig:incoherent-compcost-XBF-LAMBDA-I}}
\end{figure*}

Figure~\ref{fig:multidim-incoherent-compcost-XBF-LAMBDA-I} is a covariant view of the total computational cost density per station per voxel (sum of correlator, temporal averaging, and DFT components) for the correlator beamformer architecture taken two station parameters at a time while keeping the rest fixed at the nominal values of a \texttt{SKA-low}, \texttt{SKA-low-core}, or a \texttt{LAMBDA-I} station (grey dashed lines). The variation in the cost per voxel for $t_\textrm{acc}=1$~ms is dominated by the spatial DFT cost which is relatively constant in $D_\textrm{e}$ and $D_\textrm{s}$, and scales as $\sim N_\textrm{eps}^2$, displaying the same trends observed in Figure~\ref{fig:1D-incoherent-compcost-XBF-LAMBDA-I}. Figure~\ref{fig:multidim-incoherent-relative-compcost-XBF-LAMBDA-I} shows the ratio of the computational cost of the XBF architecture relative to the voltage beamforming architecture in Figure~\ref{fig:multidim-incoherent-compcost-BF-LAMBDA-I}. It is clear that the cost of the XBF architecture is less expensive only for $N_\textrm{eps}\lesssim 100$, and gets worse with increasing $N_\textrm{eps}$.

\subsection{Correlation and FFT (XFFT)}

In this approach, $\widetilde{V}_{a_m b_m}^{pq}$, the calibrated and inverse noise covariance weighted visibilities in station, $m$, obtained after the correlator in Equation~(\ref{eqn:intra-station-pol-visibilities}) are not weighted and beamformed using a DFT. Instead, they are gridded with weights, $\widetilde{B}_{a_m b_m}^{\alpha\beta;pq*}(\Delta\boldsymbol{r})$, which is the Fourier dual of $\widetilde{\mathcal{B}}_{a_m b_m}^{\alpha\beta;pq*}(\hat{\boldsymbol{s}}_k)$ in Equation~(\ref{eqn:intra-station-pol-xbf-img-expl}), onto a grid in the aperture plane. The gridded visibilities are Fourier transformed using an FFT rather than a DFT used in correlator beamforming,
\begin{align}
  \widetilde{\mathcal{I}}_m^{\alpha\beta}(\hat{\boldsymbol{s}}_k) &= \sum_j \delta^2 \Delta\boldsymbol{r}_j \, e^{i\frac{2\pi}{\lambda} \hat{\boldsymbol{s}}_k\cdot\Delta\boldsymbol{r}_j} \nonumber\\
  &\quad \left(\sum_{a_m,b_m} \sum_p \widetilde{B}_{a_m b_m}^{\alpha\beta;pq*}(\Delta\boldsymbol{r}_j-\Delta\boldsymbol{r}_{a_m b_m}) \, \widetilde{V}_{a_m b_m}^{pq} \right) \, . \label{eqn:intra-station-pol-img-xfft-expl}
\end{align}
The gridding and Fourier transform (using FFT) are represented by the parenthesis and outer summation, respectively. $\widetilde{B}_{a_m b_m}^{\alpha\beta;pq*}(\Delta\boldsymbol{r})$ denotes the polarimetric gridding kernel that is used to place the visibility data in polarisation state, $pq$, centred at location, $\Delta\boldsymbol{r}_{a_m b_m}$, onto a grid of locations, $\Delta\boldsymbol{r}_j$ in polarisation state, $\alpha\beta$. The purpose of $\widetilde{B}_{a_m b_m}^{\alpha\beta;pq*}(\Delta\boldsymbol{r})$ is similar \citep{Morales+2009,Bhatnagar+2008} to its counterpart gridding operator in the EPIC architecture \citep{Morales2011}, and can also incorporate the $w$-projection kernel \citep{Cornwell+2008}.

This is the mathematical counterpart to the operations performed on calibrated electric fields from the elements in EPIC, with few key differences -- (1) the calibrated visibilities can be accumulated and allowed for Earth rotation and bandwidth synthesis, (2) the gridding and spatial Fourier transform operate not on electric fields but visibilities, and (3) the spatial Fourier transform can occur on a slower cadence of $t_\textrm{acc}$.
In the radio interferometric context, $\widetilde{\mathcal{I}}_m^{\alpha\beta}(\hat{\boldsymbol{s}}_k)$ is referred to as the ``dirty image'' \citep{TMS2017,SIRA-II}. 

\begin{figure}
\includegraphics[width=\linewidth]
{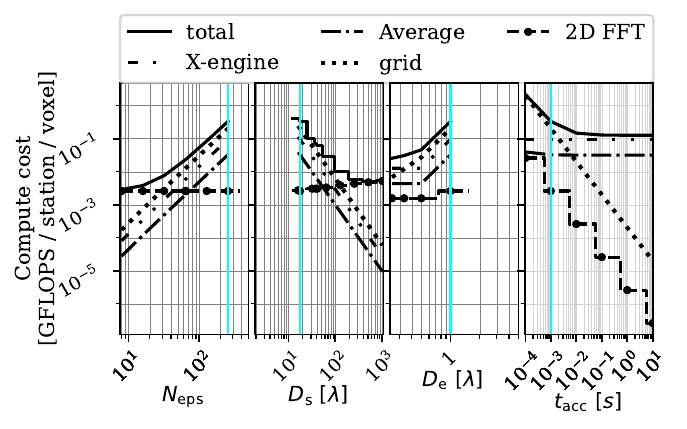}
\caption{A breakdown of the computational cost density function over the station parameters for the correlator FFT architecture (XFFT) at the station level. Each panel shows the variation with the respective parameter keeping the rest fixed at the characteristic values of the \texttt{LAMBDA-I}, \texttt{SKA-low-core}, and \texttt{SKA-low} stations (cyan lines). The correlator / X-engine (double dot- dashed lines) and gridding (dotted lines) costs are comparable and dominate over the temporal averaging (dot dashed lines) and two-dimensional FFT (dashed lines) costs for the chosen \texttt{LAMBDA-I} parameters at an imaging cadence of $t_\textrm{acc}=1$~ms.
\label{fig:1D-incoherent-compcost-XFFT-LAMBDA-I}}
\end{figure}

Figure~\ref{fig:1D-incoherent-compcost-XFFT-LAMBDA-I} shows the computational cost density breakdown for XFFT at the station level. The correlation and temporal averaging in Equation~(\ref{eqn:intra-station-pol-visibilities}) are the same as the correlator beamformer, scaling as $N_\textrm{eps}(N_\textrm{eps}-1)/2$ and have no dependence on $D_\textrm{e}$ or $D_\textrm{s}$. Hence, their costs per voxel scale as $(D_\textrm{s}/D_\textrm{e})^{-2}$. The gridding in Equation~(\ref{fig:1D-incoherent-compcost-XFFT-LAMBDA-I}) also scales as $N_\textrm{eps}(N_\textrm{eps}-1)/2$ and the size of the gridding kernel, $2^2 N_\textrm{ke}$, where, $N_\textrm{ke}$ was the size of the kernel used in EPIC. The quadrupling is because the extent of the visibility gridding kernel is expected to be twice as that of the element gridding kernel along each dimension. Thus, $2^2 N_\textrm{ke}=4$ is chosen to provide unaliased full field-of-view images for the calculations here. The incorporation of $w$-projection in the gridding kernel will correspondingly increase the kernel size, but being independent of grid dimensions, the gridding cost per voxel scales as $(D_\textrm{s}/D_\textrm{e})^{-2}$. The two-dimensional FFT in Equation~(\ref{fig:1D-incoherent-compcost-XFFT-LAMBDA-I}) depends only on the grid size and is independent of $N_\textrm{eps}$. The FFT scales as $(D_\textrm{s}/D_\textrm{e})^2\log_2(D_\textrm{s}/D_\textrm{e})^2$, and thus its cost per voxel scales only weakly as $\log_2(D_\textrm{s}/D_\textrm{e})$. The correlation and temporal averaging have to be performed at every interval of $\delta t$ and are therefore, independent of $t_\textrm{acc}$, whereas the gridding and FFT need to be performed at a cadence of $t_\textrm{acc}$, and thus scale as $t_\textrm{acc}^{-1}$. For the chosen parameters of \texttt{LAMBDA-I}, \texttt{SKA-low}, and \texttt{SKA-low-core}, and $t_\textrm{acc}=1$~ms, gridding dominates the overall cost, followed by the correlator, temporal averaging of correlations, and FFT. However, for imaging on a slower cadence, $t_\textrm{acc}\gtrsim 5-10$~ms, the cost of correlation begins to dominate. 

\begin{figure*}
\centering
\subfloat[][Covariant cost density of XFFT architecture \label{fig:multidim-incoherent-compcost-XFFT-LAMBDA-I}]
{\includegraphics[width=0.48\textwidth]
{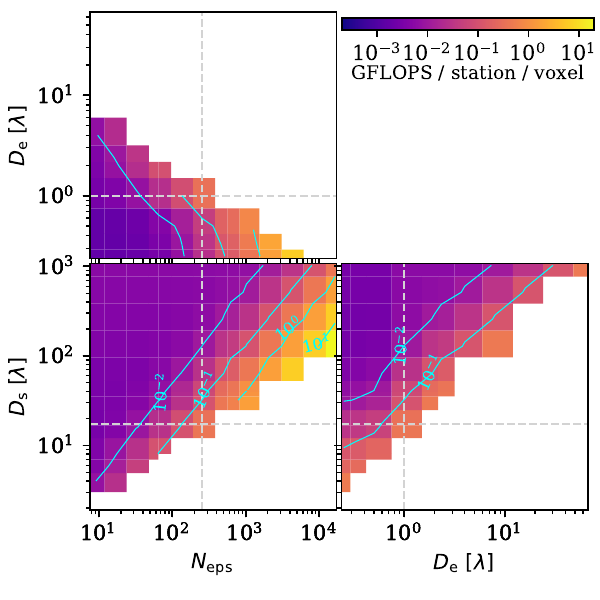}
}
\subfloat[][Covariant cost of XFFT architecture relative to BF architecture \label{fig:multidim-incoherent-relative-compcost-XFFT-LAMBDA-I}]
{\includegraphics[width=0.48\textwidth]
{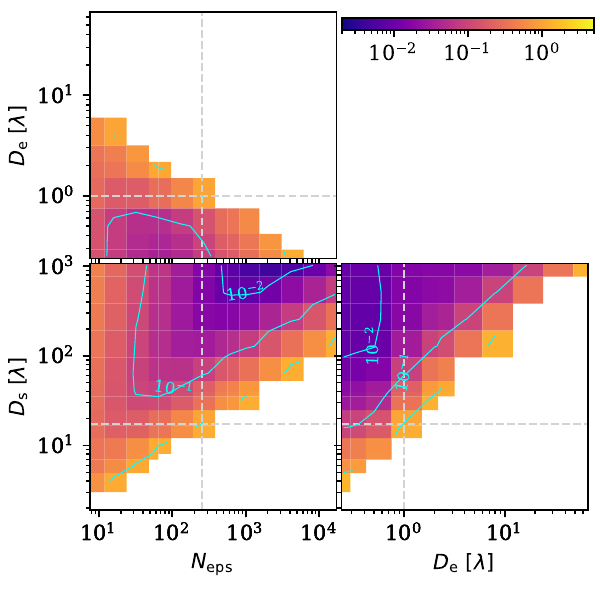}
}
\caption{(Left): Two-dimensional slices of the computational cost for imaging using an FFT of station-level cross-correlations (XFFT) for \texttt{LAMBDA-I}, \texttt{SKA-low-core}, and \texttt{SKA-low} station parameters (dashed gray lines). Cyan contours denote logarithmic levels in the colour scale. \label{fig:incoherent-compcost-XFFT-LAMBDA-I}}
\end{figure*}

Figure~\ref{fig:multidim-incoherent-compcost-XFFT-LAMBDA-I} shows a covariant view of the total computational cost density per station per voxel (sum of correlator, temporal averaging, gridding, and FFT components) for the correlator FFT architecture taken two station parameters at a time while keeping the rest fixed at the nominal values of a \texttt{LAMBDA-I}, \texttt{SKA-low}, or a \texttt{SKA-low-core} station denoted by grey dashed lines. The variation in the cost per voxel for $t_\textrm{acc}=1$~ms is dominated by the gridding cost, displaying the same trends observed in Figure~\ref{fig:1D-incoherent-compcost-XFFT-LAMBDA-I}. Figure~\ref{fig:multidim-incoherent-relative-compcost-XFFT-LAMBDA-I} shows the ratio of the computational cost of the XFFT architecture relative to the voltage beamforming architecture in Figure~\ref{fig:multidim-incoherent-compcost-BF-LAMBDA-I}. Clearly, the cost of the XFFT architecture is less expensive than the BF in most of the parameter space, particularly for larger $D_\textrm{s}/D_\textrm{e}$. For the chosen parameters (gray dashed lines), the XFFT cost roughly matches that of BF.

For applications where visibilities are processed offline and real-time imaging is not required, like traditional cosmology experiments including the EoR, only the computational cost of correlations in forming the visibilities will apply, which is still one of the dominant components of the overall cost in Figures~\ref{fig:1D-incoherent-compcost-XBF-LAMBDA-I} and \ref{fig:1D-incoherent-compcost-XFFT-LAMBDA-I}. The DFT, gridding, and FFT costs can be ignored for real-time processing.

\section{Inter-station Imaging Architectures} \label{sec:inter-station-arch}

The products from the intra-station architectures form the inputs for inter-station imaging. Here, the inter-station processing is also assumed to occur in real time. 

\subsection{Incoherent Architectures} \label{sec:incoherent}

Incoherent inter-station imaging simply involves accumulating the images from each station at a cadence of $t_\textrm{acc}$. The angular resolution remains the same as the intra-station images (assuming all stations have equal angular resolution) obtained through any of the four approaches described in section~\ref{sec:intra-station-arch}. The weighted averaging of intra-station intensities gives 
\begin{align}
    \widetilde{\mathcal{I}}^{\alpha\beta}(\hat{\boldsymbol{s}}_k) &= \frac{\sum_m w_{m}^{\alpha\beta}(\hat{\boldsymbol{s}}_k) \, \widetilde{\mathcal{I}}_m^{\alpha\beta}(\hat{\boldsymbol{s}}_k)}{\sum_m w_{m}^{\alpha\beta}(\hat{\boldsymbol{s}}_k)} \, , \label{eqn:inter-station-incoherent-pol-images}
\end{align}
where, $w_{m}^{\alpha\beta}(\hat{\boldsymbol{s}}_k)$ represents station weights for a beam pointed towards $\hat{\boldsymbol{s}}_k$. The incoherent nature of this inter-station intensity accumulation implies that it does not depend on inter-station spacing or the array diameter, $D_A$. 

\subsection{Coherent Architectures} \label{sec:coherent}

A coherent combination of data between stations requires electric field data products with phase information from the first stage of processing on individual stations. Thus, intra-station visibilities in which absolute phases have been removed and only phase differences remain, as well as the image intensity outputs of XBF and XFFT, are unusable for coherent processing at the inter-station level. Hence, the two remaining inputs are considered, namely, the electric fields, $\widetilde{\mathcal{E}}_m^\alpha(\hat{\boldsymbol{s}}_k)$, from BF and EPIC from equations~\ref{eqn:intra-station-pol-hol-img-expl} and \ref{eqn:intra-station-pol-hol-img-epic},
respectively. 

The intra-station coherent combinations, $\widetilde{\mathcal{E}}_m^\alpha(\hat{\boldsymbol{s}}_k)$, produced through BF and EPIC, respectively in Equations~(\ref{eqn:intra-station-pol-hol-img-expl}) and (\ref{eqn:intra-station-pol-hol-img-epic}), effectively convert the station, $m$, to act as virtual station-sized telescopes measuring the electric fields but simultaneously pointed towards multiple locations, denoted by $\hat{\boldsymbol{s}}_k$.
Each of these virtual station-sized telescopes has a field of view determined by inverse of the station size as $\sim (\lambda/D_\textrm{s})^2$. Together, with all the different virtual pointings, they can cover a field of view given by the inverse of the element size as $\sim (\lambda/D_\textrm{e})^2$.
Using these electric fields as the input measurements from the virtually synthesised telescopes pointed towards any given direction, $\hat{\boldsymbol{s}}_k$, the inter-station imaging architecture can now employ any of the four architectures that were previously discussed in section~\ref{sec:intra-station-arch}, for all possible directions filling the entire field of view.

For inter-station processing, all element- and station-related quantities in intra-station processing will be replaced with station- and array-related quantities, respectively. Similarly, the element-based measurements of electric fields, $\widetilde{E}_{a_m}^p$, will be replaced with the synthesised electric fields, $\widetilde{\mathcal{E}}_m^\alpha(\hat{\boldsymbol{s}}_k)$, measured by the virtual station-sized telescopes obtained either through BF (Equation~\ref{eqn:intra-station-pol-hol-img-expl}) or EPIC (Equation~\ref{eqn:intra-station-pol-hol-img-epic}).
Since the computational costs of the inter-station processing now depend on the array layout and that of intra-station processing on the station layout, it necessitates hybrid architectures appropriate for multi-scale aperture arrays. Table~\ref{tab:intra-inter-station-coherent-imaging} summarises the computational costs of inter-station processing.

\subsubsection{Beamforming}

Using the electric fields, $\widetilde{\mathcal{E}}_m^p(\hat{\boldsymbol{s}}_k)$, synthesised from each intra-station processing as input, the inter-station DFT beamforming 
can be written very similar to Equation~\ref{eqn:intra-station-pol-hol-img-expl}, as
\begin{align}
    \widetilde{\mathcal{E}}_{\hat{\boldsymbol{s}}_k}^\alpha(\boldsymbol{\sigma}_\ell) &= \sum_{m} \sum_p  \widetilde{\mathcal{W}}_{m}^{\alpha p *}(\boldsymbol{\sigma}_\ell,\hat{\boldsymbol{s}}_k) \, \widetilde{\mathcal{E}}_m^p(\hat{\boldsymbol{s}}_k) \, e^{i\frac{2\pi}{\lambda} \boldsymbol{\sigma}_\ell\cdot\boldsymbol{r}_{m}} \, . \label{eqn:inter-station-pol-hol-img-expl}   
\end{align}
$\boldsymbol{\sigma}_\ell$ denotes locations within the beam solid angle synthesised by the virtual station-sized telescopes pointed towards $\hat{\boldsymbol{s}}_k$. The directional weighting applied to the synthesised electric fields from station $m$ towards directions, $\boldsymbol{\sigma}_\ell$, centred on the beam directed towards $\hat{\boldsymbol{s}}_k$ is denoted by $\widetilde{\mathcal{W}}_{m}^{{\alpha p}^*}(\boldsymbol{\sigma}_\ell,\hat{\boldsymbol{s}}_k)$. Superscripts $\alpha$ and $p$ denote the contribution of polarisation state $p$ from the measured input field to polarisation state $\alpha$ in the synthesised field.

The polarised intensity distribution within the beam solid angle
is given by the pixel-wise outer product over polarisation states (``squaring'') and averaging, similar to Equation~(\ref{eqn:intra-station-opt-pol-img-outprod}),
\begin{align}
    \widetilde{\mathcal{I}}^{\alpha\beta}_{\hat{\boldsymbol{s}}_k}(\boldsymbol{\sigma}_\ell) &= \left\langle \widetilde{\mathcal{E}}_{\hat{\boldsymbol{s}}_k}^\alpha(\boldsymbol{\sigma}_\ell) \,  \widetilde{\mathcal{E}}_{\hat{\boldsymbol{s}}_k}^{\beta^*}(\boldsymbol{\sigma}_\ell) \right\rangle \, , \label{eqn:inter-station-opt-pol-img-outprod}
\end{align}
where, the angular brackets denote a temporal averaging across an interval of $t_\textrm{acc}$. The angular resolution of the inter-station beamforming will correspond to the dimensions of the station layout within the array as $\sim (\lambda/D_\textrm{A})^2$.

\subsubsection{EPIC}

Instead of DFT beamforming with electric fields synthesised from the stations, inter-station EPIC can be used to simultaneously form beams in all independent directions filling the available field of view using FFT. It takes the form
\begin{align}
  \widetilde{\mathcal{E}}_{\hat{\boldsymbol{s}}_k}^\alpha(\boldsymbol{\sigma}_\ell) &= \sum_j \delta^2 \boldsymbol{r}_j \, e^{i\frac{2\pi}{\lambda} \hat{\boldsymbol{\sigma}}_\ell\cdot\boldsymbol{r}_j} \nonumber\\
  &\quad \left(\sum_{m} \sum_p \widetilde{W}_{m}^{\alpha p*}(\boldsymbol{r}_j-\boldsymbol{r}_{m},\hat{\boldsymbol{s}}_k) \, \widetilde{\mathcal{E}}_m^p(\hat{\boldsymbol{s}}_k) \right) \, . \label{eqn:inter-station-pol-hol-img-epic-expl}
\end{align}
The parenthesis represents gridding of the synthesised input electric fields, $\widetilde{\mathcal{E}}_m^p(\hat{\boldsymbol{s}}_k)$ at locations, $\boldsymbol{r}_{m}$, onto the grid locations at $\boldsymbol{r}_j$ with the gridding kernels, $\widetilde{W}_{m}^{\alpha p*}(\boldsymbol{r},\hat{\boldsymbol{s}}_k)$, that are applicable for intra-station syntheses towards $\boldsymbol{s}_k$ from stations indexed by $m$. $\widetilde{W}_{m}^{\alpha p*}(\boldsymbol{r},\hat{\boldsymbol{s}}_k)$ is the Fourier dual of $\widetilde{\mathcal{W}}_{m}^{{\alpha p}^*}(\boldsymbol{\sigma}_\ell,\hat{\boldsymbol{s}}_k)$ used in DFT beamforming in Equation~(\ref{eqn:inter-station-pol-hol-img-expl}). The outer summation denotes the two-dimensional FFT of the weighted and gridded electric fields. A gridding kernel size of $N_\textrm{ks}=1^2$ is assumed. Incorporating any $w$-terms using $w$-projection into the gridding will increase $N_\textrm{ks}$. A padding of the grid can be done to control the pixel scale in the image. A constant padding factor of $\gamma_\textrm{A}=2$ (adopted in this paper) along each dimension will provide identical pixel scale and image size as that formed from gridded visibilities. The polarised intensities are produced by ``squaring'' and temporal averaging as in Equation~(\ref{eqn:inter-station-opt-pol-img-outprod}). The image within the solid angle of each intra-station beam towards $\boldsymbol{s}_k$ have a solid angle corresponding to the full array layout including all stations as $\sim (\lambda/D_\textrm{A})^2$.

For arrays that are considering FFT correlators to obtain visibilities, there will be an additional inverse-FFT (not included here) on the polarised intensities. However, the computational cost of this additional step is not expected to be significant compared to the overall cost due to the slower cadence of the operation.

\subsubsection{Correlator Beamforming}

In this correlator-based approach, the station-synthesised electric fields,
$\widetilde{\mathcal{E}}_m^p(\hat{\boldsymbol{s}}_k)$, virtually pointed towards $\hat{\boldsymbol{s}}_k$, are cross-correlated and temporally averaged to produce the inter-station visibilities, 
\begin{align}
    \widetilde{V}_{mn}^{pq}(\hat{\boldsymbol{s}}_k) &= \bigl\langle \widetilde{\mathcal{E}}_m^p(\hat{\boldsymbol{s}}_k) \, \widetilde{\mathcal{E}}_n^{q*}(\hat{\boldsymbol{s}}_k)\bigr\rangle \, . \label{eqn:inter-station-pol-visibilities}
\end{align}
The visibilities phase-centred on $\hat{\boldsymbol{s}}_k$ are then beamformed using DFT similar to Equation~(\ref{eqn:intra-station-pol-xbf-img-expl}) to produce polarised intensities at locations, $\boldsymbol{\sigma}_\ell$, within the solid angle of the beam centred on $\hat{\boldsymbol{s}}_k$ as 
\begin{align}
    \widetilde{\mathcal{I}}^{\alpha\beta}_{\hat{\boldsymbol{s}}_k}(\boldsymbol{\sigma}_\ell)
    &= \sum_{m,n} \sum_{p,q} \widetilde{\mathcal{B}}_{mn}^{\alpha\beta;pq*}(\boldsymbol{\sigma}_\ell, \hat{\boldsymbol{s}}_k) \, \widetilde{V}_{mn}^{pq}(\hat{\boldsymbol{s}}_k) \,  e^{i\frac{2\pi}{\lambda} \hat{\boldsymbol{\sigma}}_\ell\cdot\Delta\boldsymbol{r}_{m n}} \, . \label{eqn:inter-station-pol-xbf-img-expl} 
\end{align}
$\widetilde{\mathcal{B}}_{mn}^{\alpha\beta;pq*}(\boldsymbol{\sigma}_\ell,\hat{\boldsymbol{s}}_k)$ is a polarimetric directional weighting towards locations, $\boldsymbol{\sigma}_\ell$, within the solid angle of the station-synthesised beam centred on $\hat{\boldsymbol{s}}_k$, and is applied to the corresponding visibility phase-centred towards $\hat{\boldsymbol{s}}_k$. Each of the images within the beam solid angle centred on $\hat{\boldsymbol{s}}_k$ will have an angular size that scales with the array size as $\sim (\lambda/D_\textrm{A})^2$. 

\subsubsection{Correlation FFT}

An alternative to creating images from the visibilities using DFT beamforming is through FFT after gridding. The visibilities obtained in Equation~(\ref{eqn:inter-station-pol-visibilities}) are transformed using an FFT following the same formalism in Equation~(\ref{eqn:intra-station-pol-img-xfft-expl}), to yield the polarised intensities simultaneously in all independent pixels filling the field of view,
\begin{align}
  \widetilde{\mathcal{I}}_{\hat{\boldsymbol{s}}_k}^{\alpha\beta}(\boldsymbol{\sigma}_\ell) &= \sum_j \delta^2 \Delta\boldsymbol{r}_j \, e^{i\frac{2\pi}{\lambda} \boldsymbol{\sigma}_\ell\cdot\Delta\boldsymbol{r}_j} \nonumber\\
  &\quad \left(\sum_{m,n} \sum_{p,q} \widetilde{B}_{mn}^{\alpha\beta;pq*}(\Delta\boldsymbol{r}_j-\Delta\boldsymbol{r}_{mn}, \hat{\boldsymbol{s}}_k) \, \widetilde{V}_{mn}^{pq}(\hat{\boldsymbol{s}}_k) \right) \, . \label{eqn:inter-station-pol-img-xfft-expl}
\end{align}
The parenthesis represents the gridding operation. The term, $\widetilde{B}_{mn}^{\alpha\beta;pq*}(\Delta\boldsymbol{r}, \hat{\boldsymbol{s}}_k)$, is the Fourier dual of $\widetilde{\mathcal{B}}_{mn}^{\alpha\beta;pq*}(\boldsymbol{\sigma}_\ell,\hat{\boldsymbol{s}}_k)$ in Equation~(\ref{eqn:inter-station-pol-xbf-img-expl}). It acts as the polarimetric gridding kernel that weights and grids the inverse noise covariance weighted inter-station visibilities, $\widetilde{V}_{mn}^{pq}(\hat{\boldsymbol{s}}_k)$, at station separations, $\Delta\boldsymbol{r}_{mn}$, phased towards $\hat{\boldsymbol{s}}_k$, onto an array-scale grid at locations, $\Delta\boldsymbol{r}_j$. Any visibilities overlapping on the grid points are thus weighted and averaged. A gridding kernel size of $2^2 N_\textrm{ks}$ is assumed, where the quadrupling (doubling in each direction) is relative to the kernel size used in EPIC. In this paper, $2^2 N_\textrm{ks}=4$ is assumed to provide unaliased imaging within the field of view. The kernel size would increase if $w$-projection is incorporated to account for any $w$-terms across the array.

If only visibilities, not images, are required such as in traditional cosmology experiments, then the costs associated with the DFT in Equation~(\ref{eqn:inter-station-pol-xbf-img-expl}), and gridding and FFT in Equation~(\ref{eqn:inter-station-pol-img-xfft-expl}) can be ignored for real-time processing. However, the cost of correlations in forming the visibilities will continue to dominate the total cost budget for large arrays.

\begin{table*}[htb!]
\normalsize
\begin{threeparttable}
\caption{Computational budget of intra- and inter-station coherent imaging architectures.}
\label{tab:intra-inter-station-coherent-imaging}
\begin{tabular}{cccccc}
\toprule
\headrow Architecture & Components & Intra-station & Inter-station & Timescale & Equations \\
 & & FLOP count & FLOP count & & \\ 
 & & (per station & (per voxel) & & \\
 & & per voxel) & & & \\
\midrule\midrule
BF & Beamforming & $8 n_\textrm{p}^2 N_\textrm{eps}$ & $8 n_\textrm{p}^2 N_\textrm{s}$ & $\delta t$ & \ref{eqn:intra-station-pol-hol-img-expl}, \ref{eqn:inter-station-pol-hol-img-expl} \\
& Squaring & $6 n_\textrm{p}^2$ & $6 n_\textrm{p}^2$ & $\delta t$ & \ref{eqn:intra-station-opt-pol-img-outprod}, \ref{eqn:inter-station-opt-pol-img-outprod}  \\
& Accumulation & $2 n_\textrm{p}^2$ & $2 n_\textrm{p}^2$ & $\delta t$ & \ref{eqn:intra-station-opt-pol-img-outprod}, \ref{eqn:inter-station-opt-pol-img-outprod} \\
\midrule
EPIC & Gridding\tnote{a} & $\frac{6 n_\textrm{p}^2 N_\textrm{eps} N_\textrm{ke}}{(D_\textrm{s}/D_\textrm{e})^2}$ & $\frac{6 n_\textrm{p}^2 N_\textrm{s} N_\textrm{ks}}{(D_\textrm{A}/D_\textrm{s})^2}$ & $\delta t$ & \ref{eqn:intra-station-pol-hol-img-epic}, \ref{eqn:inter-station-pol-hol-img-epic-expl}   \\
& FFT\tnote{b}$\,\,\,$\tnote{c} & $8 n_\textrm{p} \frac{r}{\log_2 r} \gamma_\textrm{s}^2 \log_2\left(\gamma_\textrm{s}\frac{D_\textrm{s}}{D_\textrm{e}}\right)^2$ & $8 n_\textrm{p} \frac{r}{\log_2 r} \gamma_\textrm{A}^2 \log_2\left(\gamma_\textrm{A}\frac{D_\textrm{A}}{D_\textrm{s}}\right)^2$ & $\delta t$ & \ref{eqn:intra-station-pol-hol-img-epic}, \ref{eqn:inter-station-pol-hol-img-epic-expl}   \\
& Squaring & $6 n_\textrm{p}^2 \gamma_\textrm{s}^2$ & $6 n_\textrm{p}^2 \gamma_\textrm{A}^2$ & $\delta t$ & \ref{eqn:intra-station-opt-pol-img-outprod}, \ref{eqn:inter-station-opt-pol-img-outprod}  \\ 
& Accumulation & $2 n_\textrm{p}^2 \gamma_\textrm{s}^2$ & $2 n_\textrm{p}^2 \gamma_\textrm{A}^2$ & $\delta t$ & \ref{eqn:intra-station-opt-pol-img-outprod}, \ref{eqn:inter-station-opt-pol-img-outprod}  \\
\midrule
XBF & Correlator & $\frac{6 n_\textrm{p}^2 N_\textrm{eps} (N_\textrm{eps}-1)/2}{(D_\textrm{s}/D_\textrm{e})^2}$ & $\frac{6 n_\textrm{p}^2 N_\textrm{s} (N_\textrm{s}-1)/2}{(D_\textrm{A}/D_\textrm{s})^2}$ & $\delta t$ & \ref{eqn:intra-station-pol-visibilities}, \ref{eqn:inter-station-pol-visibilities}  \\
& Accumulation & $\frac{2 n_\textrm{p}^2 N_\textrm{eps} (N_\textrm{eps}-1)/2}{(D_\textrm{s}/D_\textrm{e})^2}$ & $\frac{2 n_\textrm{p}^2  N_\textrm{s} (N_\textrm{s}-1)/2}{(D_\textrm{A}/D_\textrm{s})^2}$ & $\delta t$ & \ref{eqn:intra-station-pol-visibilities}, \ref{eqn:inter-station-pol-visibilities}  \\
& DFT & $8 n_\textrm{p}^2 N_\textrm{eps} (N_\textrm{eps}-1)/2$ & $8 n_\textrm{p}^2 N_\textrm{s} (N_\textrm{s}-1)/2$ & $t_\textrm{acc}$ & \ref{eqn:intra-station-pol-xbf-img-expl}, \ref{eqn:inter-station-pol-xbf-img-expl}  \\
\midrule
XFFT & Correlator & $\frac{6 n_\textrm{p}^2 N_\textrm{eps} (N_\textrm{eps}-1)/2}{(D_\textrm{s}/D_\textrm{e})^2}$ & $\frac{6 n_\textrm{p}^2 N_\textrm{s} (N_\textrm{s}-1)/2}{(D_\textrm{A}/D_\textrm{s})^2}$ & $\delta t$ & \ref{eqn:intra-station-pol-visibilities}, \ref{eqn:inter-station-pol-visibilities}  \\
& Accumulation & $\frac{2 n_\textrm{p}^2 N_\textrm{eps} (N_\textrm{eps}-1)/2}{(D_\textrm{s}/D_\textrm{e})^2}$ & $\frac{2 n_\textrm{p}^2 N_\textrm{s} (N_\textrm{s}-1)/2}{(D_\textrm{A}/D_\textrm{s})^2}$ & $\delta t$ & \ref{eqn:intra-station-pol-visibilities}, \ref{eqn:inter-station-pol-visibilities} \\
& Gridding\tnote{a} & $\frac{8 n_\textrm{p}^4 (2^2 N_\textrm{ke}) N_\textrm{eps} (N_\textrm{eps}-1)/2}{(D_\textrm{s}/D_\textrm{e})^2}$ & $\frac{8 n_\textrm{p}^4 (2^2 N_\textrm{ks}) N_\textrm{s} (N_\textrm{s}-1)/2}{(D_\textrm{A}/D_\textrm{s})^2}$ & $t_\textrm{acc}$ & \ref{eqn:intra-station-pol-img-xfft-expl}, \ref{eqn:inter-station-pol-img-xfft-expl} \\
& FFT\tnote{b} & $8 n_\textrm{p}^2 \frac{r}{\log_2 r} \log_2\left(\frac{D_\textrm{s}}{D_\textrm{e}}\right)^2$ & $8 n_\textrm{p}^2 \frac{r}{\log_2 r} \log_2\left(\frac{D_\textrm{A}}{D_\textrm{s}}\right)^2$ & $t_\textrm{acc}$ & \ref{eqn:intra-station-pol-img-xfft-expl}, \ref{eqn:inter-station-pol-img-xfft-expl} \\
\bottomrule
\end{tabular}
\begin{tablenotes}[hang]
\item[a]$N_\textrm{ke}$ and $N_\textrm{ks}$ denote the sizes (in number of cells) of the gridding kernels of the antenna elements and the stations, respectively. In this paper, $N_\textrm{ke}=1^2$ and $N_\textrm{ks}=1^2$. For XFFT, compared to pre-correlated data used in EPIC, the gridding sizes for intra- and inter-station correlations are quadrupled to $2^2 N_\textrm{ke}$ and $2^2 N_\textrm{ks}$, respectively to account for the expansion of the correlated kernels and to remove the aliasing effects within the respective fields of view. If the gridding kernels incorporate $w$-projection, the kernel sizes will be correspondingly larger.
\item[b]$r$ denotes the radix of the FFT algorithm \citep{Cooley+Tukey1965}. Here, $r=2$. 
\item[c]$\gamma_\textrm{s}$ and $\gamma_\textrm{A}$ denote padding factors for station- and array-level FFT, respectively in EPIC. Here, $\gamma_\textrm{s}=2$ and $\gamma_\textrm{A}=2$ along each dimension to achieve identical image and pixel sizes as that from XFFT.
\end{tablenotes}
\end{threeparttable}
\end{table*}

\section{Results \& Discussion} \label{sec:results}

Below is a comparative view of the various architectures from the viewpoint of computational cost density for the range of array layouts and imaging cadence considered in this paper. 

\subsection{Station-scale Imaging Architectures}\label{sec:station-scale-imaging}

\begin{figure}
\centering
\subfloat[][Station imaging compute cost densities for \texttt{LAMBDA-I}, \texttt{SKA-low}, and \texttt{SKA-low-core} stations. \label{fig:1D-incoherent-compcost-LAMBDA-I}]
{\includegraphics[width=\textwidth]
{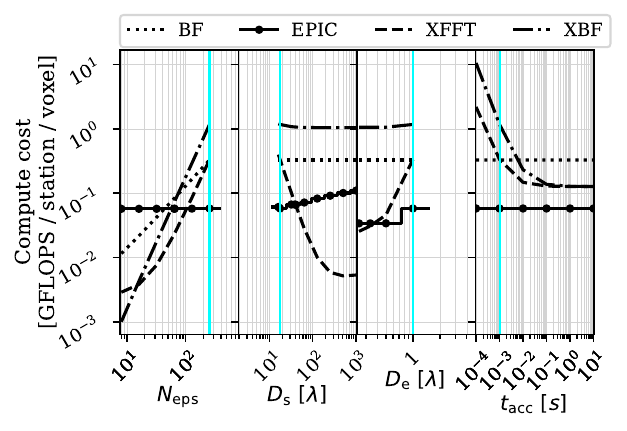}
} \\
\subfloat[][Station imaging compute cost densities for a \texttt{CASPA} station.\label{fig:1D-incoherent-compcost-CASPA}]
{\includegraphics[width=\textwidth]
{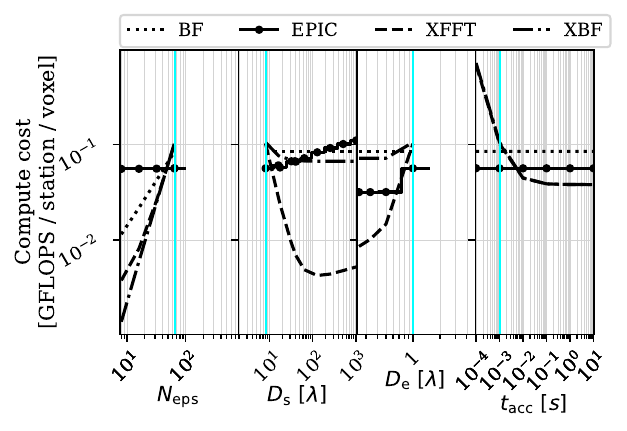}
} \\
\subfloat[][Station imaging compute cost densities for a \texttt{FarView-core} station. \label{fig:1D-incoherent-compcost-FarView}]
{\includegraphics[width=\textwidth]
{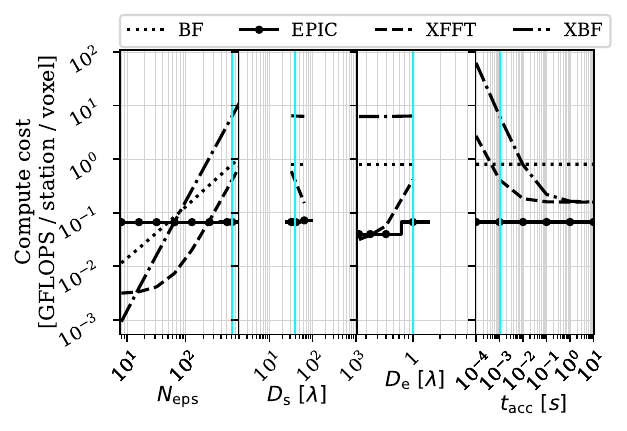}
}
\caption{Computational cost densities of different station-level imaging architectures for (a) \texttt{SKA-low-core}, \texttt{SKA-low}, and \texttt{LAMBDA-I}, (b) \texttt{CASPA}, and (c) \texttt{FarView-core} stations. In each subpanel, all parameters are fixed at the array's nominal value (cyan line) except the parameter shown on the $x$-axis. For \texttt{SKA-low-core}, \texttt{SKA-low}, \texttt{LAMBDA-I}, and \texttt{FarView-core}, the EPIC architecture is found to be most efficient computationally. For \texttt{CASPA}, EPIC is more efficient than other architectures when imaging cadence is $t_\textrm{acc}\lesssim 1-5$~ms, while XFFT is more efficient for $t_\textrm{acc}\gtrsim 5-10$~ms.
  \label{fig:1D-incoherent-compcost}}
\end{figure}

Figure~\ref{fig:1D-incoherent-compcost} compares the computational cost densities of different imaging architectures discussed in section~\ref{sec:intra-station-arch} for varying parameters of the arrays in Table~\ref{tab:array_params}. The station layout parameters for \texttt{LAMBDA-I}, \texttt{SKA-low-core}, and \texttt{SKA-low} are identical, and so are their station-level computing costs. For station parameters of these arrays and that of \texttt{FarView-core}, EPIC is computationally advantageous clearly. For \texttt{CASPA}, EPIC and XFFT hold the advantage for imaging cadence faster and slower than $t_\textrm{acc}\simeq 5$~ms, respectively.

\subsection{Multi-scale Imaging Architectures}\label{sec:multi-scale-imaging}

Figure~\ref{fig:1D-coherent-compcost-a} shows a comparison of the compute cost densities in the discovery phase space of different multi-scale imaging architectures combining four options -- BF, EPIC, XBF, and XFFT -- on the inter-station data with two options -- BF and EPIC -- on the intra-station data for various array parameters and imaging cadences.  

\begin{figure*}
\centering
\subfloat[][\texttt{LAMBDA-I} \label{fig:1D-coherent-compcost-LAMBDA-I}]
{\includegraphics[width=\textwidth]
{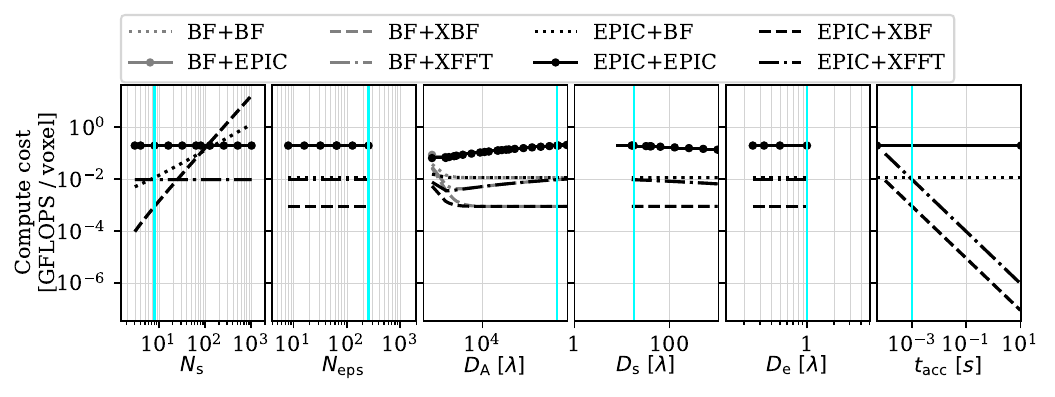}
} \\
\subfloat[][\texttt{SKA-Low-core} \label{fig:1D-coherent-compcost-SKA-low-core}]
{\includegraphics[width=\textwidth]
{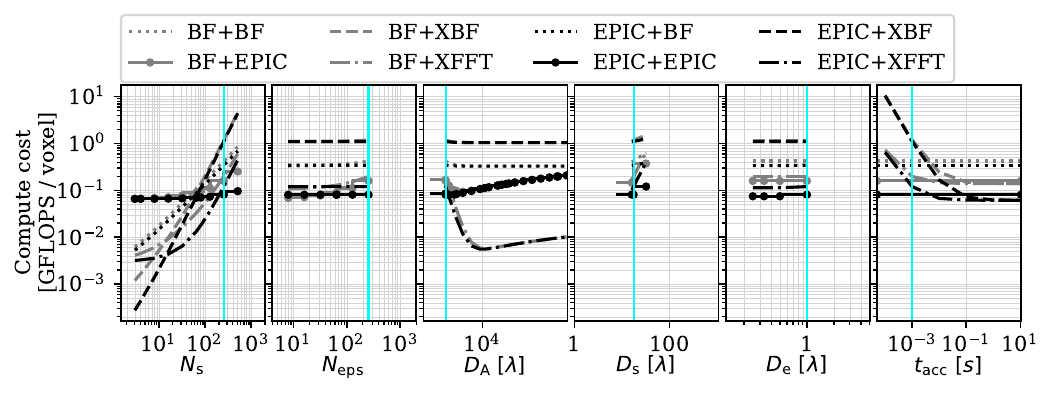}
} \\
\subfloat[][\texttt{SKA-Low} \label{fig:1D-coherent-compcost-SKA-low}]
{\includegraphics[width=\textwidth]
{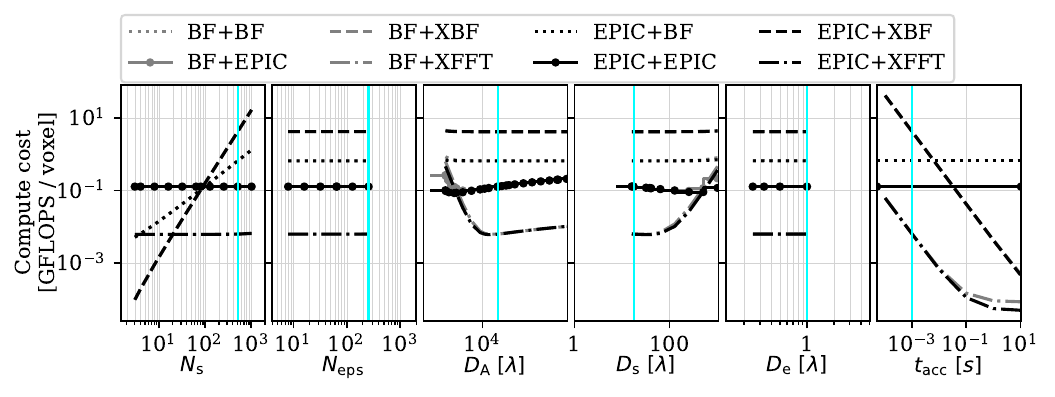}
} 
\caption{One-dimensional slices of the computational cost density for (a) \texttt{LAMBDA-I}, (b) \texttt{SKA-low-core}, and (c) \texttt{SKA-low} using two-stage coherent imaging at intra- and inter-station levels. Station-synthesised electric fields are obtained through BF (gray) and EPIC (black) in intra-station processing. Inter-station processing of data is obtained by BF (dotted), EPIC (solid), XBF (dashed), and XFFT (dot-dashed). The text in the legend lists the first (intra-station) and second (inter-station) stage architectures delimited by `+'. \label{fig:1D-coherent-compcost-a}}
\end{figure*}

\begin{figure*}\ContinuedFloat
\subfloat[][\texttt{CASPA} \label{fig:1D-coherent-compcost-CASPA}]
{\includegraphics[width=\textwidth]
{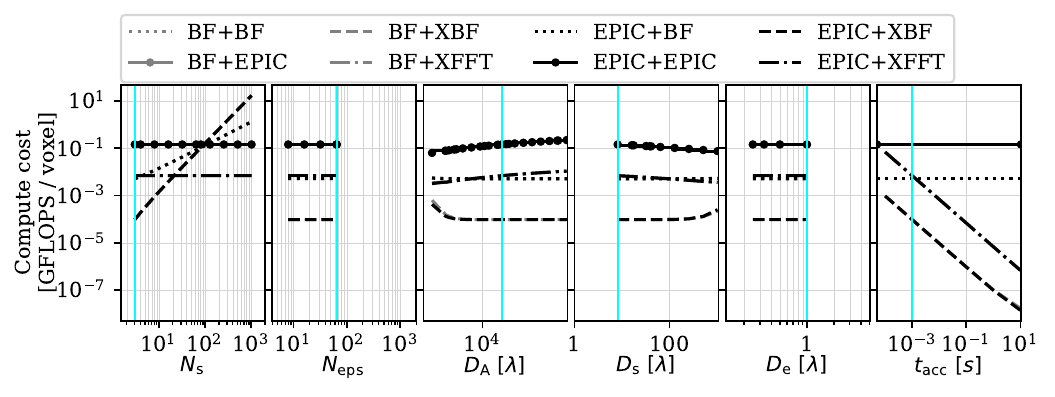}
} \\
\subfloat[][\texttt{FarView-core} \label{fig:1D-coherent-compcost-FarView}]
{\includegraphics[width=\textwidth]
{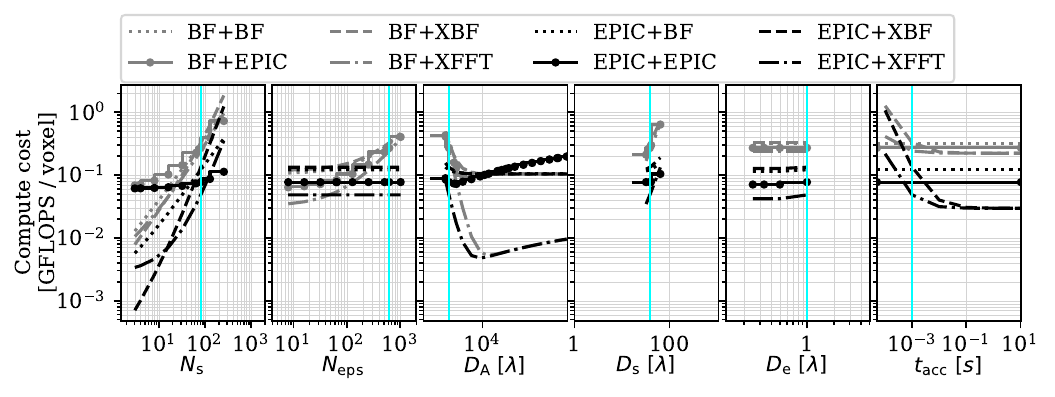}
}
\caption{Contd. Same as above but for (d) \texttt{CASPA}, and (e) \texttt{FarView-core}. \label{fig:1D-coherent-compcost-b}}
\end{figure*}

In all cases, EPIC (black lines) is more efficient than BF (gray lines) for intra-station full field of view data processing (first stage). This is because of the relatively large number of elements densely packed in stations and due to the capability of simultaneous full field of view beamforming by the EPIC architecture powered by the FFT's efficiency. 

The inter-station processing, however, varies significantly due to the diversity in the various inter-station array parameters and imaging cadence considered. For example, 
\begin{itemize}
    \item \texttt{LAMBDA-I} will benefit from deploying the correlator beamforming (XBF) architecture for inter-station processing by a factor of $\simeq 10$ over XFFT or BF. 
    \item \texttt{SKA-low-core} will have a marginal advantage from EPIC for the chosen parameters. However, if the number of stations, $N_\textrm{s}$, decreases,  the array diameter, $D_\textrm{A}$, increases or if the imaging cadence interval is slowed to $t_\textrm{acc} \gtrsim 5$~ms, the XFFT architecture starts to have a significant advantage over EPIC on the inter-station scale of data processing. 
    \item \texttt{SKA-low} will have an advantage from the XFFT architecture for inter-station processing by a factor $\simeq 20$.
    \item \texttt{CASPA} will have an advantage by a factor $\gtrsim 20$ from using the XBF architecture for inter-station processing due to significantly smaller number of stations.
    \item \texttt{FarView-core} appears to have a marginal advantage in using XFFT for inter-station processing for the chosen parameters. If $N_\textrm{s}$ increases or if the cadence interval decreases to $t_\textrm{acc}\lesssim 1$~ms, then EPIC becomes more efficient. 
\end{itemize}

\subsection{Effect of Imaging Cadence}\label{sec:cadence}

The scientific goals enabled by access to the wide range of timescales exhibited by transient phenomena (microseconds to days) are key determinants of whether the architecture deployed can support the processing, as they are intricately tied to different timescales (see Table~\ref{tab:intra-inter-station-coherent-imaging}). Here, the impact of the cadence interval on the choice of imaging architecture for multi-scale arrays for intra- and inter-station processing is examined. Table~\ref{tab:cadence} summarises the findings from Figures~\ref{fig:1D-incoherent-compcost} and \ref{fig:1D-coherent-compcost-b} based on the dependence of the computational cost densities of the different architectures and arrays on $t_\textrm{acc}$.

\begin{table*}[htb!]
\normalsize
\begin{threeparttable}
\caption{Impact of imaging cadence on choice of efficient imaging architecture.}
\label{tab:cadence}
\begin{tabular}{c|cccc|cccc}
\toprule
\headrow 
Array & \multicolumn{4}{c|}{Intra-station architecture} & \multicolumn{4}{c}{Inter-station architecture} \\ 
 & $\lesssim 0.1$~ms & $\simeq 1$~ms & $\gtrsim 100$~ms & $\gtrsim 10$~s & $\lesssim 0.1$~ms & $\simeq 1$~ms & $\gtrsim 100$~ms & $\gtrsim 10$~s \\ 
\midrule\midrule

\texttt{LAMBDA-I} & EPIC & EPIC & EPIC & EPIC & BF/XBF & XBF & XBF & XBF \\
\midrule
\texttt{SKA-low-core} & EPIC & EPIC & EPIC & EPIC & EPIC & EPIC & XFFT/EPIC/XBF & XBF/XFFT \\
\midrule
\texttt{SKA-low} & EPIC & EPIC & EPIC & EPIC & XFFT/EPIC & XFFT & XFFT & XFFT \\
\midrule
\texttt{CASPA} & EPIC & EPIC & XFFT & XFFT & XBF & XBF & XBF & XBF \\
\midrule
\texttt{FarView-core} & EPIC & EPIC & EPIC & EPIC & EPIC & XFFT & XBF/XFFT & XBF/XFFT \\
\bottomrule
\end{tabular}
\end{threeparttable}
\end{table*}

For intra-station processing, EPIC emerges as the most efficient among various architectures for all arrays considered on all timescales by a factor of $\lesssim 10$, except for \texttt{CASPA} when $t_\textrm{acc}\gtrsim 100$~ms where the XFFT architecture emerges to be more efficient than EPIC by a factor of few. This is because the stations have a sufficiently large number of elements packed densely (from Table~\ref{tab:array_params}). For arrays like \texttt{CASPA} that have lesser number of elements and/or are not as densely packed, the cadence interval becomes an important factor. For cadence intervals, $t_\textrm{acc}\gtrsim 100$~ms, the XFFT architecture becomes advantageous. Even for cosmological applications that only require intra-station visibilities at cadence slower than 10~s, an FFT-correlator architecture implemented via EPIC will have the computational advantage over a traditional correlator architecture. However, for CASPA, the latter will have an advantage over the former.

The same reasoning can be extended for inter-station processing, where there is significant diversity. There is a large range in the number of stations and in the array filling factor. As a result, both the layout and cadence interval play significant roles in determining the efficiency of architectures. For arrays with small numbers of stations such as \texttt{LAMBDA-I} and \texttt{CASPA}, the XBF architecture holds an advantage by a factor of $\simeq 10$ on $t_\textrm{acc}\gtrsim 1$~ms and BF for $t_\textrm{acc}\lesssim 0.1$~ms. For arrays with large numbers of stations such as \texttt{SKA-low-core} with a reasonably high packing density, EPIC holds the advantage on all timescales by a factor of few, but with slower cadence intervals, $t_\textrm{acc}\gtrsim 100$~ms, XBF and XFFT also have comparable efficiency. Although the \texttt{SKA-low} has a large number of stations, they are not as densely packed as in the core, and therefore, the role of $t_\textrm{acc}$ is more significant. XFFT and EPIC are comparable (within a factor of few) at $t_\textrm{acc}\lesssim 1$~ms, but for $t_\textrm{acc}\gtrsim 1$~ms, XFFT is clearly efficient by a factor $\gtrsim 10$. Similarly, \texttt{FarView-core} has a comparable array filling factor as \texttt{SKA-low-core}, but has far fewer stations. Hence, EPIC is only marginally efficient for $t_\textrm{acc}\lesssim 0.1$~ms, whereas XFFT becomes more efficient by a factor of few for $t_\textrm{acc}\gtrsim 1$~ms. If cosmological applications require inter-station visibilities at a cadence slower than $\gtrsim 100$~ms, a correlator-based architecture holds the computational edge for all the arrays considered here.

\subsection{Other considerations}\label{sec:other-metrics}

In transient search and cosmological applications, there will be additional processing required. For example, a de-dispersion trial process is required to discover new FRBs and pulsars with optimum signal-to-noise ratio. Computationally efficient algorithms for de-dispersion like the Fast Dispersion Measure Transform \citep[FDMT;][]{Zackay+2014} need to be performed either on intensities in the image plane pixels or on visibilities in the aperture plane, the cost of which will scale with the number of independent pixels in the image or the visibilities, respectively. For compact, dense, large-$N$ arrays, the number of independent pixels will be typically lesser than the number of measured visibilities, and thus it will be advantageous to perform de-dispersion in the image plane. Conversely, it will be computationally preferable to de-disperse visibilities for sparse arrays. A full accounting of such costs for most general cases is beyond the scope of this paper.

In intensity mapping applications for cosmology, the images need to be post-processed for deconvolution, foreground, removal, etc. As these post-processing computations are additional and common to all architectures, including them in adopted cost metric will not affect the results. Calibration is another aspect that is not included in this computational cost analysis. Because calibration typically operates on a significantly slower cadence compared to $t_\textrm{acc}$, and the computational cost is not tedious \citep{Beardsley+2017,Gorthi+2021}, its contribution is negligible to the overall cost budget, and therefore will not affect the results.

Besides computational costs, there are other practical considerations in the deployment of an architecture such as the hardware's processing bandwidth, internal memory bandwidth management, input and output data rates, etc. For example, the GPU-based deployment of EPIC on the LWA is severely limited by the memory bandwidth constraints on chip\footnote{See \url{https://github.com/epic-astronomy/Memos/blob/master/PDFs/009_EPIC_Code_Optimizations.md} for a detailed analysis.}. The architecture of current generation of GPUs is found to be inefficient at handling the transpose operation required in a two-dimensional FFT, whereas they may be more efficient at handling correlations. On the contrary, the current generation of FPGAs seem more suited for low bit width two-dimensional FFT from computational and on-chip memory bandwidth perspectives. It is extremely difficult to forecast the continually evolving landscape of hardware capabilities accessible to instruments at the instant of deployment. Thus, this analysis has been restricted to the metric of computational cost density, which is independent of evolving hardware capabilities. For example, even if the computations are parallelised over a large number of GPU cores as noted in \citet{Sokolowski+2024}, the total computational cost remains unaffected. So, it is intended to be viewed as a first step among many towards an actual implementation. 

\section{Conclusions} \label{sec:conclusion}

Explosive transient phenomena and observations of cosmological structures are two high impact science drivers in radio astronomy. The paradigm of aperture arrays used in radio astronomy is undergoing a transformation towards large numbers of low-cost elements grouped together in a hierarchy of spatial scales to achieve the desired science outcomes. Such a paradigm faces enormous challenges in computing and throughput because of the large numbers of elements spanning a multiplicity of spatial scales, a range of cadence intervals, real-time processing, and a continuous operation model. Different processing architectures are optimal for specific layouts and processing cadences. However, a single architectural may not be optimal for a hierarchical spatial layout.

This paper explores hybrid architecture solutions for existing and planned multi-scale aperture arrays (\texttt{SKA-low-core}, \texttt{SKA-low}, \texttt{LAMBDA-I}, \texttt{CASPA}, and \texttt{FarView-core}) on a wide range of cadence intervals, 0.1~ms$\le t_\textrm{acc} \lesssim 10$~s, over the full field of view, using the metric of computational cost density evaluated over the discovery parameter space. The interferometric data processing architectures considered here are pre-correlation voltage beamformer (BF), a generic FFT-based direct imager (EPIC), beamforming from correlations (XBF), and FFT-based imaging of correlations (XFFT). The compute budget for various architecture combinations for these different arrays are broken down into their detailed components. 

For densely packed layouts with large numbers of elements, EPIC is computationally quite efficient regardless of processing cadence. This was the case for the station layouts of all the arrays considered, and hence EPIC emerges as the computationally most efficient architecture for full field of view, station-level data processing on all cadence intervals except for \texttt{CASPA} on $t_\textrm{acc}\gtrsim 100$~ms, when XFFT becomes more efficient, although marginally. For cosmological applications with large stations where only intra-station visibilities are required at a slower cadence $\gtrsim 10$~s with the arrays studied here, an EPIC-based FFT correlator holds the advantage over a traditional correlator architecture, except in the case of CASPA where the latter will be preferred.

The inter-station array layout has a wider range in the number of stations and density of their distribution. In such cases, the cadence interval plays a significant role in determining the computationally most efficient architecture. For inter-station data processing in \texttt{LAMBDA-I} and \texttt{CASPA}, XBF is found to be the most suitable option for all cadence intervals, while for \texttt{SKA-low-core} it is EPIC that holds the advantage on all cadence intervals. For \texttt{SKA-low}, XFFT appears most efficient for $t_\textrm{acc}\gtrsim 1$~ms, while EPIC and XFFT become comparable for $t_\textrm{acc}\lesssim 0.1$~ms. If only slower cadence ($\gtrsim 10$~s) inter-station visibilities are required from these arrays such as in cosmological applications, a traditional correlator architecture holds a computational advantage. Thus, a multi-scale aperture array requires an optimal hybrid architecture strategy depending on the processing cadence, which in turn depends on the scientific objectives. This study provides a guide for designing computationally strategic data processing architecture for hierarchical, multi-scale aperture arrays, and conversely, can be used to design array parameters for a specified computational capability. 

This work considers only computational cost density over the discovery parameter space as a metric for finding the most efficient combination of processing architectures for multi-scale aperture arrays. Additional computing will be required for post-processing steps such as de-dispersion, deconvolution, etc. depending on the application. Actual deployment will have to take other significant practical factors into account such as power consumption, memory bandwidth, data rate, cost and capabilities of current hardware.

\begin{acknowledgement}
Inputs from Ron Ekers, Vivek Gupta, David Humphrey, Nivedita Mahesh, and Rajaram Nityananda are gratefully acknowledged. 
\end{acknowledgement}







\printendnotes


\begin{thebibliography}{}
\expandafter\ifx\csname natexlab\endcsname\relax\def\natexlab#1{#1}\fi

\bibitem[{{Beardsley} {et~al.}(2017){Beardsley}, {Thyagarajan}, {Bowman}, \&
  {Morales}}]{Beardsley+2017}
{Beardsley}, A.~P., {Thyagarajan}, N., {Bowman}, J.~D., \& {Morales}, M.~F.
  2017, \mnras, 470, 4720

\bibitem[{{Bhatnagar} {et~al.}(2008){Bhatnagar}, {Cornwell}, {Golap}, \&
  {Uson}}]{Bhatnagar+2008}
{Bhatnagar}, S., {Cornwell}, T.~J., {Golap}, K., \& {Uson}, J.~M. 2008, \aap,
  487, 419

\bibitem[{{Bochenek} {et~al.}(2020){Bochenek}, {Ravi}, {Belov}, {Hallinan},
  {Kocz}, {Kulkarni}, \& {McKenna}}]{Bochenek+2020}
{Bochenek}, C.~D., {Ravi}, V., {Belov}, K.~V., {et~al.} 2020, \nat, 587, 59

\bibitem[{{Braun} {et~al.}(2019){Braun}, {Bonaldi}, {Bourke}, {Keane}, \&
  {Wagg}}]{SKA1+2019}
{Braun}, R., {Bonaldi}, A., {Bourke}, T., {Keane}, E., \& {Wagg}, J. 2019,
  arXiv e-prints, arXiv:1912.12699

\bibitem[{{Chandra} {et~al.}(2016){Chandra}, {Anupama}, {Arun}, {Iyyani},
  {Misra}, {Narasimha}, {Ray}, {Resmi}, {Roy}, \& {Sutaria}}]{Chandra+2016}
{Chandra}, P., {Anupama}, G.~C., {Arun}, K.~G., {et~al.} 2016, Journal of
  Astrophysics and Astronomy, 37, 30

\bibitem[{Cooley \& Tukey(1965)}]{Cooley+Tukey1965}
Cooley, J.~W., \& Tukey, J.~W. 1965, Mathematics of Computation, 19, 297

\bibitem[{{Cordes}(2007)}]{Cordes2007}
{Cordes}, J. 2007, in From Planets to Dark Energy: the Modern Radio Universe,
  35

\bibitem[{{Cornwell} {et~al.}(2008){Cornwell}, {Golap}, \&
  {Bhatnagar}}]{Cornwell+2008}
{Cornwell}, T.~J., {Golap}, K., \& {Bhatnagar}, S. 2008, IEEE Journal of
  Selected Topics in Signal Processing, 2, 647

\bibitem[{{Cosmic Visions 21 cm Collaboration} {et~al.}(2018){Cosmic Visions 21
  cm Collaboration}, {Ansari}, {Arena}, {Bandura}, {Bull}, {Castorina},
  {Chang}, {Chen}, {Connor}, {Foreman}, {Frisch}, {Green}, {Johnson},
  {Karagiannis}, {Liu}, {Masui}, {Meerburg}, {M{\"u}nchmeyer}, {Newburgh},
  {Obuljen}, {O'Connor}, {Padmanabhan}, {Shaw}, {Sheehy}, {Slosar}, {Smith},
  {Stankus}, {Stebbins}, {Timbie}, {Villaescusa-Navarro}, {Wallisch}, \&
  {White}}]{CosmicVisions+2018}
{Cosmic Visions 21 cm Collaboration}, {Ansari}, R., {Arena}, E.~J., {et~al.}
  2018, arXiv e-prints, arXiv:1810.09572

\bibitem[{{Crawford} {et~al.}(2022){Crawford}, {Hisano}, {Golden}, {Kikunaga},
  {Laity}, \& {Zoeller}}]{Crawford+2022}
{Crawford}, F., {Hisano}, S., {Golden}, M., {et~al.} 2022, \mnras, 515, 3698

\bibitem[{{Crichton} {et~al.}(2022){Crichton}, {Aich}, {Amara}, {Bandura},
  {Bassett}, {Bengaly}, {Berner}, {Bhatporia}, {Bucher}, {Chang}, {Chiang},
  {Cliche}, {Crichton}, {Dave}, {De Villiers}, {Dobbs}, {Ewall-Wice}, {Eyono},
  {Finlay}, {Gaddam}, {Ganga}, {Gayley}, {Gerodias}, {Gibbon}, {Gumba},
  {Gupta}, {Harris}, {Heilgendorff}, {Hilton}, {Hincks}, {Hitz}, {Jalilvand},
  {Julie}, {Kader}, {Kania}, {Karagiannis}, {Karastergiou}, {Kesebonye},
  {Kittiwisit}, {Kneib}, {Knowles}, {Kuhn}, {Kunz}, {Maartens}, {MacKay},
  {MacPherson}, {Monstein}, {Moodley}, {Mugundhan}, {Naidoo}, {Naidu},
  {Newburgh}, {Nistane}, {Di Nitto}, {{\"O}l{\c{c}}ek}, {Pan}, {Paul},
  {Peterson}, {Pieters}, {Pieterse}, {Pillay}, {Polish}, {Randrianjanahary},
  {Refregier}, {Renard}, {Retana-Montenegro}, {Rout}, {Russeeawon}, {Sadr},
  {Saliwanchik}, {Sampath}, {Sanghavi}, {Santos}, {Sengate}, {Shaw}, {Sievers},
  {Smirnov}, {Smith}, {Sob}, {Srianand}, {Stronkhorst}, {Sunder},
  {Tartakovsky}, {Taylor}, {Timbie}, {Tolley}, {Townsend}, {Tyndall},
  {Ungerer}, {van Dyk}, {van Vuuren}, {Vanderlinde}, {Viant}, {Walters},
  {Wang}, {Weltman}, {Woudt}, {Wulf}, {Zavyalov}, \& {Zhang}}]{HIRAX+2022}
{Crichton}, D., {Aich}, M., {Amara}, A., {et~al.} 2022, Journal of Astronomical
  Telescopes, Instruments, and Systems, 8, 011019

\bibitem[{{Daishido} {et~al.}(1991){Daishido}, {Asuma}, {Nishibori},
  {Nakajima}, {Yano}, {Otobe}, {Watanabe}, {Tsuchiya}, \&
  {Iwase}}]{Daishido+1991}
{Daishido}, T., {Asuma}, K., {Nishibori}, K., {et~al.} 1991, in Astronomical
  Society of the Pacific Conference Series, Vol.~19, IAU Colloq. 131: Radio
  Interferometry. Theory, Techniques, and Applications, ed. T.~J. {Cornwell} \&
  R.~A. {Perley}, 86--89

\bibitem[{{DeBoer} {et~al.}(2017){DeBoer}, {Parsons}, {Aguirre}, {Alexander},
  {Ali}, {Beardsley}, {Bernardi}, {Bowman}, {Bradley}, {Carilli}, {Cheng}, {de
  Lera Acedo}, {Dillon}, {Ewall-Wice}, {Fadana}, {Fagnoni}, {Fritz},
  {Furlanetto}, {Glendenning}, {Greig}, {Grobbelaar}, {Hazelton}, {Hewitt},
  {Hickish}, {Jacobs}, {Julius}, {Kariseb}, {Kohn}, {Lekalake}, {Liu}, {Loots},
  {MacMahon}, {Malan}, {Malgas}, {Maree}, {Martinot}, {Mathison}, {Matsetela},
  {Mesinger}, {Morales}, {Neben}, {Patra}, {Pieterse}, {Pober}, {Razavi-Ghods},
  {Ringuette}, {Robnett}, {Rosie}, {Sell}, {Smith}, {Syce}, {Tegmark},
  {Thyagarajan}, {Williams}, \& {Zheng}}]{HERA+2017}
{DeBoer}, D.~R., {Parsons}, A.~R., {Aguirre}, J.~E., {et~al.} 2017, \pasp, 129,
  045001

\bibitem[{{Dewdney} {et~al.}(2009){Dewdney}, {Hall}, {Schilizzi}, \&
  {Lazio}}]{Dewdney+2009}
{Dewdney}, P.~E., {Hall}, P.~J., {Schilizzi}, R.~T., \& {Lazio}, T.~J.~L.~W.
  2009, IEEE Proceedings, 97, 1482

\bibitem[{{Dowell} \& {Taylor}(2018)}]{Dowell+2018}
{Dowell}, J., \& {Taylor}, G.~B. 2018, Journal of Astronomical Instrumentation,
  7, 1850006

\bibitem[{{Du} {et~al.}(2009){Du}, {Xu}, {Qiao}, \& {Han}}]{Du+2009}
{Du}, Y.~J., {Xu}, R.~X., {Qiao}, G.~J., \& {Han}, J.~L. 2009, \mnras, 399,
  1587

\bibitem[{{Eilek} \& {Hankins}(2016)}]{Eilek+2016}
{Eilek}, J.~A., \& {Hankins}, T.~H. 2016, Journal of Plasma Physics, 82,
  635820302

\bibitem[{{Foster} {et~al.}(2014){Foster}, {Hickish}, {Magro}, {Price}, \&
  {Zarb Adami}}]{Foster+2014}
{Foster}, G., {Hickish}, J., {Magro}, A., {Price}, D., \& {Zarb Adami}, K.
  2014, \mnras, 439, 3180

\bibitem[{{Gorthi} {et~al.}(2021){Gorthi}, {Parsons}, \&
  {Dillon}}]{Gorthi+2021}
{Gorthi}, D.~B., {Parsons}, A.~R., \& {Dillon}, J.~S. 2021, \mnras, 500, 66

\bibitem[{{Gupta} {et~al.}(2022){Gupta}, {Flynn}, {Farah}, {Bailes}, {Deller},
  {Day}, \& {Lower}}]{Gupta+2022}
{Gupta}, V., {Flynn}, C., {Farah}, W., {et~al.} 2022, \mnras, 514, 5866

\bibitem[{{Hankins} \& {Eilek}(2007)}]{Hankins+2007}
{Hankins}, T.~H., \& {Eilek}, J.~A. 2007, \apj, 670, 693

\bibitem[{{Hankins} {et~al.}(2003){Hankins}, {Kern}, {Weatherall}, \&
  {Eilek}}]{Hankins+2003}
{Hankins}, T.~H., {Kern}, J.~S., {Weatherall}, J.~C., \& {Eilek}, J.~A. 2003,
  \nat, 422, 141

\bibitem[{{Haskell} {et~al.}(2018){Haskell}, {Zdunik}, {Fortin}, {Bejger},
  {Wijnands}, \& {Patruno}}]{Haskell+2018}
{Haskell}, B., {Zdunik}, J.~L., {Fortin}, M., {et~al.} 2018, \aap, 620, A69

\bibitem[{{Keane}(2013)}]{Keane2013}
{Keane}, E.~F. 2013, in Neutron Stars and Pulsars: Challenges and Opportunities
  after 80 years, ed. J.~{van Leeuwen}, Vol. 291, 295--300

\bibitem[{{Kent} {et~al.}(2019){Kent}, {Dowell}, {Beardsley}, {Thyagarajan},
  {Taylor}, \& {Bowman}}]{Kent+2019}
{Kent}, J., {Dowell}, J., {Beardsley}, A., {et~al.} 2019, \mnras, 486, 5052

\bibitem[{{Kent} {et~al.}(2020){Kent}, {Beardsley}, {Bester}, {Gull},
  {Nikolic}, {Dowell}, {Thyagarajan}, {Taylor}, \& {Bowman}}]{Kent+2020}
{Kent}, J., {Beardsley}, A.~P., {Bester}, L., {et~al.} 2020, \mnras, 491, 254

\bibitem[{{Krishnan} {et~al.}(2023){Krishnan}, {Beardsley}, {Bowman}, {Dowell},
  {Kolopanis}, {Taylor}, \& {Thyagarajan}}]{Krishnan+2023}
{Krishnan}, H., {Beardsley}, A.~P., {Bowman}, J.~D., {et~al.} 2023, \mnras,
  520, 1928

\bibitem[{{Lorimer} {et~al.}(2007){Lorimer}, {Bailes}, {McLaughlin},
  {Narkevic}, \& {Crawford}}]{Lorimer+2007}
{Lorimer}, D.~R., {Bailes}, M., {McLaughlin}, M.~A., {Narkevic}, D.~J., \&
  {Crawford}, F. 2007, Science, 318, 777

\bibitem[{{Luo} {et~al.}(2024){Luo}, {Ekers}, {Hobbs}, {Dunning}, {James},
  {Lower}, {Gupta}, {Zic}, {Sokolowski}, {Phillips}, {Deller}, \&
  {Staveley-Smith}}]{Luo+2024}
{Luo}, R., {Ekers}, R.~D., {Hobbs}, G., {et~al.} 2024, arXiv e-prints,
  arXiv:2405.07439

\bibitem[{{Masui} {et~al.}(2019){Masui}, {Shaw}, {Ng}, {Smith}, {Vanderlinde},
  \& {Paradise}}]{Masui+2019}
{Masui}, K.~W., {Shaw}, J.~R., {Ng}, C., {et~al.} 2019, \apj, 879, 16

\bibitem[{{McLaughlin} {et~al.}(2006){McLaughlin}, {Lyne}, {Lorimer}, {Kramer},
  {Faulkner}, {Manchester}, {Cordes}, {Camilo}, {Possenti}, {Stairs}, {Hobbs},
  {D'Amico}, {Burgay}, \& {O'Brien}}]{Mclaughlin+2006}
{McLaughlin}, M.~A., {Lyne}, A.~G., {Lorimer}, D.~R., {et~al.} 2006, \nat, 439,
  817

\bibitem[{{Morales}(2011)}]{Morales2011}
{Morales}, M.~F. 2011, \pasp, 123, 1265

\bibitem[{{Morales} \& {Matejek}(2009)}]{Morales+2009}
{Morales}, M.~F., \& {Matejek}, M. 2009, \mnras, 400, 1814

\bibitem[{{Morales} \& {Wyithe}(2010)}]{Morales+2010}
{Morales}, M.~F., \& {Wyithe}, J. S.~B. 2010, \araa, 48, 127

\bibitem[{{Nimmo} {et~al.}(2022){Nimmo}, {Hessels}, {Kirsten}, {Keimpema},
  {Cordes}, {Snelders}, {Hewitt}, {Karuppusamy}, {Archibald}, {Bezrukovs},
  {Bhardwaj}, {Blaauw}, {Buttaccio}, {Cassanelli}, {Conway}, {Corongiu},
  {Feiler}, {Fonseca}, {Forss{\'e}n}, {Gawro{\'n}ski}, {Giroletti}, {Kharinov},
  {Leung}, {Lindqvist}, {Maccaferri}, {Marcote}, {Masui}, {Mckinven},
  {Melnikov}, {Michilli}, {Mikhailov}, {Ng}, {Orbidans}, {Ould-Boukattine},
  {Paragi}, {Pearlman}, {Petroff}, {Rahman}, {Scholz}, {Shin}, {Smith},
  {Stairs}, {Surcis}, {Tendulkar}, {Vlemmings}, {Wang}, {Yang}, \&
  {Yuan}}]{Nimmo+2022}
{Nimmo}, K., {Hessels}, J.~W.~T., {Kirsten}, F., {et~al.} 2022, Nature
  Astronomy, 6, 393

\bibitem[{{Otobe} {et~al.}(1994){Otobe}, {Nakajima}, {Nishibori}, {Saito},
  {Kobayashi}, {Tanaka}, {Watanabe}, {Aramaki}, {Hoshikawa}, {Asuma}, \&
  {Daishido}}]{Otobe+1994}
{Otobe}, E., {Nakajima}, J., {Nishibori}, K., {et~al.} 1994, \pasj, 46, 503

\bibitem[{{Philippov} {et~al.}(2019){Philippov}, {Uzdensky}, {Spitkovsky}, \&
  {Cerutti}}]{Philippov+2019}
{Philippov}, A., {Uzdensky}, D.~A., {Spitkovsky}, A., \& {Cerutti}, B. 2019,
  \apjl, 876, L6

\bibitem[{{Pietka} {et~al.}(2015){Pietka}, {Fender}, \& {Keane}}]{Pietka+2015}
{Pietka}, M., {Fender}, R.~P., \& {Keane}, E.~F. 2015, \mnras, 446, 3687

\bibitem[{{Polidan} {et~al.}(2024){Polidan}, {Burns}, {Ignatiev}, {Hegedus},
  {Pober}, {Mahesh}, {Chang}, {Hallinan}, {Ning}, \& {Bowman}}]{Polidan+2024}
{Polidan}, R.~S., {Burns}, J.~O., {Ignatiev}, A., {et~al.} 2024, Advances in
  Space Research, 74, 528

\bibitem[{{Price}(2024)}]{Price2024}
{Price}, D.~C. 2024, \pasa, 41, e037

\bibitem[{{Pritchard} \& {Loeb}(2012)}]{Pritchard+2012}
{Pritchard}, J.~R., \& {Loeb}, A. 2012, Reports on Progress in Physics, 75,
  086901

\bibitem[{Reddy {et~al.}(2024)Reddy, Bowman, Dowell, Taylor, Beardsley, \&
  Taylor}]{Reddy+2024}
Reddy, K., Bowman, J.~D., Dowell, J., {et~al.} 2024, in Software and
  Cyberinfrastructure for Astronomy VIII, ed. J.~Ibsen \& G.~Chiozzi, Vol.
  13101, International Society for Optics and Photonics (SPIE), 131011C

\bibitem[{Slosar {et~al.}(2019)Slosar, Ahmed, Alonso, Amin, Arena, Bandura,
  Battaglia, Blazek, Bull, Castorina, Chang, Connor, Dav{\' e}, Dvorkin,
  Engelen, Ferraro, Flauger, Foreman, Frisch, Green, Holder, Jacobs, Johnson,
  Dillon, Karagiannis, Kaurov, Knox, Liu, Loverde, Ma, Masui, McClintock,
  Moodley, Munchmeyer, Newburgh, Ng, Nomerotski, O\textquoteright{}Connor,
  Obuljen, Padmanabhan, Parkinson, Prochaska, Rajendran, Rapetti, Saliwanchik,
  Schaan, Sehgal, Shaw, Sheehy, Sheldon, Shirley, Silverstein, Slatyer, Slosar,
  Stankus, Stebbins, Timbie, Tucker, Tyndall, Navarro, Wallisch, \&
  White}]{PUMA+2019}
Slosar, A., Ahmed, Z., Alonso, D., {et~al.} 2019, Bulletin of the AAS, 51,
  https://baas.aas.org/pub/2020n7i053

\bibitem[{{Snelders} {et~al.}(2023){Snelders}, {Nimmo}, {Hessels}, {Bensellam},
  {Zwaan}, {Chawla}, {Ould-Boukattine}, {Kirsten}, {Faber}, \&
  {Gajjar}}]{Snelders+2023}
{Snelders}, M.~P., {Nimmo}, K., {Hessels}, J.~W.~T., {et~al.} 2023, Nature
  Astronomy, 7, 1486

\bibitem[{{Sokolowski} {et~al.}(2024){Sokolowski}, {Aniruddha}, {Di
  Pietrantonio}, {Harris}, {Price}, {McSweeney}, {Wayth}, \&
  {Bhat}}]{Sokolowski+2024}
{Sokolowski}, M., {Aniruddha}, G., {Di Pietrantonio}, C., {et~al.} 2024, arXiv
  e-prints, arXiv:2405.13478

\bibitem[{{Taylor} {et~al.}(1999){Taylor}, {Carilli}, \& {Perley}}]{SIRA-II}
{Taylor}, G.~B., {Carilli}, C.~L., \& {Perley}, R.~A., eds. 1999, Astronomical
  Society of the Pacific Conference Series, Vol. 180, {Synthesis Imaging in
  Radio Astronomy II} (San Francisco, Calif. : Astronomical Society of the
  Pacific)

\bibitem[{Taylor {et~al.}(2012)Taylor, Ellingson, Kassim, Craig, Dowell, Wolfe,
  Hartman, Bernardi, Clarke, Cohen, Dalal, Erickson, Hicks, Greenhill, Jacoby,
  Lane, Lazio, Mitchell, Navarro, Ord, Pihlstr\"{o}m, Polisensky, Ray, Rickard,
  Schinzel, Schmitt, Sigman, Soriano, Stewart, Stovall, Tremblay, Wang, Weiler,
  White, \& Wood}]{Taylor+2012}
Taylor, G.~B., Ellingson, S.~W., Kassim, N.~E., {et~al.} 2012, Journal of
  Astronomical Instrumentation, 01, 1250004

\bibitem[{{Tegmark}(1997)}]{Tegmark1997a}
{Tegmark}, M. 1997, \apjl, 480, L87

\bibitem[{{Tegmark} \& {Zaldarriaga}(2009)}]{Tegmark+2009}
{Tegmark}, M., \& {Zaldarriaga}, M. 2009, \prd, 79, 083530

\bibitem[{{Tegmark} \& {Zaldarriaga}(2010)}]{Tegmark+2010}
---. 2010, \prd, 82, 103501

\bibitem[{{Thompson} {et~al.}(2017){Thompson}, {Moran}, \& {Swenson}}]{TMS2017}
{Thompson}, A.~R., {Moran}, J.~M., \& {Swenson}, George~W., J. 2017,
  {Interferometry and Synthesis in Radio Astronomy, 3rd Edition} (Springer,
  Cham), doi:10.1007/978-3-319-44431-4

\bibitem[{{Thornton} {et~al.}(2013){Thornton}, {Stappers}, {Bailes},
  {Barsdell}, {Bates}, {Bhat}, {Burgay}, {Burke-Spolaor}, {Champion}, {Coster},
  {D'Amico}, {Jameson}, {Johnston}, {Keith}, {Kramer}, {Levin}, {Milia}, {Ng},
  {Possenti}, \& {van Straten}}]{Thornton+2013}
{Thornton}, D., {Stappers}, B., {Bailes}, M., {et~al.} 2013, Science, 341, 53

\bibitem[{{Thyagarajan} {et~al.}(2019){Thyagarajan}, {Beardsley}, {Bowman},
  {Dowell}, {Taylor}, {Kent}, \& {Jacobs}}]{Thyagarajan+2019}
{Thyagarajan}, N., {Beardsley}, A., {Bowman}, J., {et~al.} 2019, in Bulletin of
  the American Astronomical Society, Vol.~51, 263

\bibitem[{{Thyagarajan} {et~al.}(2017){Thyagarajan}, {Beardsley}, {Bowman}, \&
  {Morales}}]{Thyagarajan+2017}
{Thyagarajan}, N., {Beardsley}, A.~P., {Bowman}, J.~D., \& {Morales}, M.~F.
  2017, \mnras, 467, 715

\bibitem[{{Thyagarajan} {et~al.}(2011){Thyagarajan}, {Helfand}, {White}, \&
  {Becker}}]{Thyagarajan+2011}
{Thyagarajan}, N., {Helfand}, D.~J., {White}, R.~L., \& {Becker}, R.~H. 2011,
  \apj, 742, 49

\bibitem[{{Tingay} {et~al.}(2013){Tingay}, {Goeke}, {Bowman}, {Emrich}, {Ord},
  {Mitchell}, {Morales}, {Booler}, {Crosse}, {Wayth}, {Lonsdale}, {Tremblay},
  {Pallot}, {Colegate}, {Wicenec}, {Kudryavtseva}, {Arcus}, {Barnes},
  {Bernardi}, {Briggs}, {Burns}, {Bunton}, {Cappallo}, {Corey}, {Deshpande},
  {Desouza}, {Gaensler}, {Greenhill}, {Hall}, {Hazelton}, {Herne}, {Hewitt},
  {Johnston-Hollitt}, {Kaplan}, {Kasper}, {Kincaid}, {Koenig}, {Kratzenberg},
  {Lynch}, {Mckinley}, {Mcwhirter}, {Morgan}, {Oberoi}, {Pathikulangara},
  {Prabu}, {Remillard}, {Rogers}, {Roshi}, {Salah}, {Sault}, {Udaya-Shankar},
  {Schlagenhaufer}, {Srivani}, {Stevens}, {Subrahmanyan}, {Waterson},
  {Webster}, {Whitney}, {Williams}, {Williams}, \& {Wyithe}}]{Tingay+2013}
{Tingay}, S.~J., {Goeke}, R., {Bowman}, J.~D., {et~al.} 2013, \pasa, 30, e007

\bibitem[{{van Haarlem} {et~al.}(2013){van Haarlem}, {Wise}, {Gunst}, {Heald},
  {McKean}, {Hessels}, {de Bruyn}, {Nijboer}, {Swinbank}, {Fallows},
  {Brentjens}, {Nelles}, {Beck}, {Falcke}, {Fender}, {H{\"o}randel},
  {Koopmans}, {Mann}, {Miley}, {R{\"o}ttgering}, {Stappers}, {Wijers},
  {Zaroubi}, {van den Akker}, {Alexov}, {Anderson}, {Anderson}, {van Ardenne},
  {Arts}, {Asgekar}, {Avruch}, {Batejat}, {B{\"a}hren}, {Bell}, {Bell}, {van
  Bemmel}, {Bennema}, {Bentum}, {Bernardi}, {Best}, {B{\^\i}rzan}, {Bonafede},
  {Boonstra}, {Braun}, {Bregman}, {Breitling}, {van de Brink}, {Broderick},
  {Broekema}, {Brouw}, {Br{\"u}ggen}, {Butcher}, {van Cappellen}, {Ciardi},
  {Coenen}, {Conway}, {Coolen}, {Corstanje}, {Damstra}, {Davies}, {Deller},
  {Dettmar}, {van Diepen}, {Dijkstra}, {Donker}, {Doorduin}, {Dromer}, {Drost},
  {van Duin}, {Eisl{\"o}ffel}, {van Enst}, {Ferrari}, {Frieswijk}, {Gankema},
  {Garrett}, {de Gasperin}, {Gerbers}, {de Geus}, {Grie{\ss}meier}, {Grit},
  {Gruppen}, {Hamaker}, {Hassall}, {Hoeft}, {Holties}, {Horneffer}, {van der
  Horst}, {van Houwelingen}, {Huijgen}, {Iacobelli}, {Intema}, {Jackson},
  {Jelic}, {de Jong}, {Juette}, {Kant}, {Karastergiou}, {Koers}, {Kollen},
  {Kondratiev}, {Kooistra}, {Koopman}, {Koster}, {Kuniyoshi}, {Kramer},
  {Kuper}, {Lambropoulos}, {Law}, {van Leeuwen}, {Lemaitre}, {Loose}, {Maat},
  {Macario}, {Markoff}, {Masters}, {McFadden}, {McKay-Bukowski}, {Meijering},
  {Meulman}, {Mevius}, {Middelberg}, {Millenaar}, {Miller-Jones}, {Mohan},
  {Mol}, {Morawietz}, {Morganti}, {Mulcahy}, {Mulder}, {Munk}, {Nieuwenhuis},
  {van Nieuwpoort}, {Noordam}, {Norden}, {Noutsos}, {Offringa}, {Olofsson},
  {Omar}, {Orr{\'u}}, {Overeem}, {Paas}, {Pandey-Pommier}, {Pandey}, {Pizzo},
  {Polatidis}, {Rafferty}, {Rawlings}, {Reich}, {de Reijer}, {Reitsma},
  {Renting}, {Riemers}, {Rol}, {Romein}, {Roosjen}, {Ruiter}, {Scaife}, {van
  der Schaaf}, {Scheers}, {Schellart}, {Schoenmakers}, {Schoonderbeek},
  {Serylak}, {Shulevski}, {Sluman}, {Smirnov}, {Sobey}, {Spreeuw}, {Steinmetz},
  {Sterks}, {Stiepel}, {Stuurwold}, {Tagger}, {Tang}, {Tasse}, {Thomas},
  {Thoudam}, {Toribio}, {van der Tol}, {Usov}, {van Veelen}, {van der Veen},
  {ter Veen}, {Verbiest}, {Vermeulen}, {Vermaas}, {Vocks}, {Vogt}, {de Vos},
  {van der Wal}, {van Weeren}, {Weggemans}, {Weltevrede}, {White}, {Wijnholds},
  {Wilhelmsson}, {Wucknitz}, {Yatawatta}, {Zarka}, {Zensus}, \& {van
  Zwieten}}]{vanHaarlem+2013}
{van Haarlem}, M.~P., {Wise}, M.~W., {Gunst}, A.~W., {et~al.} 2013, \aap, 556,
  A2

\bibitem[{{Zackay} \& {Ofek}(2017)}]{Zackay+2014}
{Zackay}, B., \& {Ofek}, E.~O. 2017, \apj, 835, 11

\bibitem[{{Zhang} {et~al.}(2023){Zhang}, {Tian}, {Zarka}, {Louis}, {Lu}, {Gao},
  {Sun}, {Yu}, {Chen}, {Cheng}, \& {Wang}}]{Zhang+2023}
{Zhang}, J., {Tian}, H., {Zarka}, P., {et~al.} 2023, \apj, 953, 65

\end{thebibliography}

\end{document}